\documentclass[twocolumn]{openjournal}

\usepackage{lipsum}

\usepackage{xcolor}
\usepackage{textgreek}
\usepackage[utf8]{inputenc}
\usepackage[english]{babel}

\usepackage{hyperref}
\hypersetup{
    unicode, 
    colorlinks=true,
    linkcolor=linkcolor,
    citecolor=linkcolor,
    filecolor=linkcolor,
    urlcolor=linkcolor,
}
\usepackage{color,colortbl}
\definecolor{linkcolor}{rgb}{0.0,0.3,0.5}
\usepackage{tensind}
\tensordelimiter{?}
\DeclareGraphicsExtensions{.bmp,.png,.jpg,.pdf}
\usepackage{verbatim}
\usepackage[normalem]{ulem}
\usepackage{orcidlink}
\usepackage{soul}

\graphicspath{ {./figs/} }

\begin{document}
\title[Ionizing radiation of {[}WR{]} stars]{Ionizing radiation of [WR] stars: atmospheres and photoionization models}

\author{Da Conceição, L. V.\orcidlink{0000-0002-5042-443X}}
\email{daconcel@myumanitoba.ca}
\affiliation{Department of Physics \& Astronomy, University of Manitoba, Winnipeg, MB R3T 2N2, Canada}

\author{Marcolino, W. L. F.\orcidlink{0000-0002-2808-6008}}
\email{wagner@ov.br}
\affiliation{Observatório do Valongo, Universidade Federal do Rio de Janeiro, Ladeira Pedro Antônio, 43, CEP 20080-090, Rio de Janeiro, Brazil}

\author{Maria, V. C.\orcidlink{0009-0009-4115-9388}}
\email{vitorcm@ufrj.br}
\affiliation{Observatório do Valongo, Universidade Federal do Rio de Janeiro, Ladeira Pedro Antônio, 43, CEP 20080-090, Rio de Janeiro, Brazil}

\begin{abstract}
The radiation field of central stars of planetary nebulae (CSPN) plays a fundamental role in shaping the origin and physical conditions of planetary nebulae (PNe). Many studies in the literature model PNe using either the blackbody approximation (bb) or, more realistically, plane-parallel (p-p) atmospheres as the ionizing source. However, these approaches become inconsistent when the central star is a Wolf-Rayet ([WR]) type. These objects are hydrogen-deficient and exhibit intense stellar winds, similar to classical massive Wolf-Rayet stars (e.g., $\sim$10$^{-7}$–10$^{-6}$ M$_\odot$ year$^{-1}$). Their accurate description therefore requires sophisticated NLTE expanding atmosphere models. In this work, we selected [WR] atmosphere models from a grid spanning a range of temperatures and mass-loss rates, and compared them with equivalent models using the bb and p-p approximations. We examined differences in the spectral energy distribution and in the ionizing photon fluxes for H I, He I, and He II, aiming to constrain the influence of stellar winds on the interaction between a nebula and its central star. In addition, using the photoionization code \texttt{CLOUDY}, we explored the impact of different types of ionizing sources on the prediction of various line ratios (I$_{\lambda}$/I$_{\mathrm{H}\beta}$) for a sample of real PNe with [WR] central stars (NGC 5315, NGC 6905, NGC 2867, NGC 40, and BD+303639). To evaluate the model performance, we used the mean root squared (rms) difference between the observed and predicted line ratios, where the best fit corresponded to the lowest rms value. For temperatures below $\sim$100,000 K, our results show that the shape of the EUV spectrum differs significantly among the p-p, bb, and [WR] cases. The stellar wind has a dramatic effect on the EUV flux, introducing substantial opacity compared with no-wind models. The $\log Q(\mathrm{He,II})$ values can decrease by several orders of magnitude. For hotter [WR] stars, the wind's effect on the number of ionizing photons is considerably reduced. For the nebulae in our sample hosting late-type [WR] central stars, photoionization models that include the stellar wind provide much better agreement with the observed line ratios than the bb and p-p approximations. Conversely, for early-type stars, the choice of ionizing source appears to play only a minor role. While a more detailed nebular analysis for each object is warranted, our findings strongly suggest that the bb and p-p approximations should be avoided when modelling PNe with [WR] central stars, particularly those of the [WCL] subtype.
\end{abstract}

\begin{keywords}
    {Planetary Nebulae, [WR] stars, Central stars of planetary nebulae}
\end{keywords}

\maketitle

\section{Introduction}
\label{sec:sec1}

Planetary Nebulae (PNe) represent the final evolutionary stages of low- and intermediate-mass stars and play a key role in enriching the interstellar medium with elements synthesized during earlier phases of stellar evolution. Beyond their chemical contribution, PNe serve as natural laboratories for testing astrophysical theories that cannot be directly verified on Earth, providing valuable insights into stellar physics and the broader processes governing stellar evolution.

According to canonical models of stellar evolution (e.g., \citealt{iben1983asymptotic}), these objects host hydrogen-rich central stars and, at this stage, are expected to evolve towards the white dwarf cooling track in the H--R diagram. However, approximately 25--30\% of the known PNe contain hydrogen-deficient central stars (\citealt{acker2003quantitative,weidmann2011central}), a class of intriguing objects whose chemical composition cannot be explained by standard evolutionary models. The most widely accepted explanation for the hydrogen deficiency in these stars is the \textit{born-again} scenario, in which the star experiences a late re-ignition of the helium-burning shell, causing it to return temporarily to the asymptotic giant branch (AGB) in the H--R diagram. Nevertheless, alternative mechanisms have been proposed. For instance, \cite{fujimoto1977origin} suggested that a very late thermal pulse (VLTP) could account for the observed lack of hydrogen. In this case, the star undergoes a VLTP while already on the white dwarf cooling track; the helium-flash convection zone engulfs the remaining envelope, and the hydrogen in the outer layers is mixed downward and completely burned. \cite{herwig2001internal} proposed another possible mechanism, in which a final thermal pulse during the AGB phase, combined with dredge-up, can lead to hydrogen depletion if the remaining envelope mass is sufficiently low. This scenario is known as the asymptotic giant branch final thermal pulse (AFTP). An additional possibility is the late thermal pulse (LTP), which occurs when the star is in the post-AGB phase, evolving at approximately constant luminosity (see \citealt{blocker1995stellar}).

One of the most intriguing types of hydrogen-deficient CSPNe are the [WR] stars. These objects share spectral characteristics with the classical massive Wolf-Rayet stars, as their spectra are dominated by strong emission lines produced by powerful stellar winds (\citealt{acker2003quantitative,marcolino2007detailed,mendez1991photospheric}). In terms of chemical composition, [WR] stars exhibit primarily helium, carbon, oxygen, and neon, elements synthesized during helium burning and subsequent evolutionary phases (\citealt{werner2006elemental}). 

A large fraction of hydrogen-deficient CSPNe are classified as \textbf{[WC]}-type. This designation refers to a subclass within the broader Wolf-Rayet ([WR]) category, where the letter ‘C’ denotes spectra dominated by strong \textbf{carbon} and helium emission lines. The \textbf{square brackets~[]} are used to distinguish these \textbf{low-mass} objects (central stars of planetary nebulae) from their \textbf{massive} counterparts, as both groups display remarkably similar spectral features.

In terms of temperature, these stars are generally divided into two groups: the late-type ([WCL]) stars, with surface temperatures between 20 and 80~kK (\citealt{leuenhagen1996spectral,leuenhagen1998spectral}), and the early-type ([WCE]) stars, with temperatures ranging from 80 to 150~kK (\citealt{koesterke1997,pena1998}). Along with the [WC] stars, other objects belong to the class of hydrogen-deficient CSPNe. For instance, the [WC] descendants, the PG~1159 stars, exhibit weaker stellar winds compared to [WC] stars, with mass-loss rates of $\dot{M} \sim 10^{-7}$--$10^{-8}$~M$_{\odot}$~yr$^{-1}$ (\citealt{koesterke1998mass}). Another example is the small group of extremely hot post-AGB stars, the O(He) type (T~$\gtrsim$~100,000~K). These objects display nearly pure He~\textsc{ii} line spectra and can have even weaker winds, with mass-loss rates estimated at $\dot{M} \sim 10^{-7}$--$10^{-10}$~M$_{\odot}$~yr$^{-1}$ (see, e.g., \citealt{rauch1998spectral}). 

The other two chemically distinct subtypes of [WR] CSPNe are the nitrogen ([WN]) and oxygen ([WO]) stars. They are characterized by broad and intense emission lines, primarily of helium, nitrogen, carbon, and oxygen, formed in their dense stellar winds. [WN] stars show strong He~\textsc{ii} and N~\textsc{v} lines, indicative of high surface temperatures and nitrogen enrichment, while [WO] stars exhibit prominent O~\textsc{vi} and C~\textsc{iv} lines, reflecting even more extreme surface conditions and later evolutionary stages (e.g., \citealt{acker2003quantitative,todt2010central}). 

Regarding PNe, the central star provides the radiation field that powers the nebula, producing the emission observed from Earth through photoionization processes. To study the physics of these environments, a common approach is to construct photoionization models that reproduce the observed spectra. In this context, most studies employ plane-parallel stellar atmosphere models to represent the ionizing source in their photoionization calculations (\citealt{bohigas2008photoionization,miller2016analysis,rubio2022planetary}), and in some cases, even simpler approximations such as a blackbody model (e.g., \citealt{howard1997detailed,henry2015co}). However, when the central stars are of the [WR] type, such approximations become inconsistent, as they neglect the effects of the dense stellar wind on the emergent flux, which is crucial for powering the surrounding nebula.

\cite{rauch2003grid} performed a differential analysis of the ionizing fluxes predicted by stellar atmosphere models for CSPNe. They compared spectral energy distributions (SEDs) from NLTE plane-parallel models with various chemical compositions and temperature ranges against corresponding blackbody SEDs. Their results showed that even for simple models containing only H and He, the flux level at the He~\textsc{ii} ionization edge differed dramatically from that of a blackbody (see their Fig.~1). The analysis demonstrated that in the ultraviolet region, the main driver of photoionization in PNe (see the review by \citealt{morisset2016photoionization}), it would be necessary to combine multiple blackbody components to approximate the UV characteristics of plane-parallel models. 

In this paper, we extend this analysis by incorporating expanding atmosphere models into the comparison. Furthermore, we investigate how stellar winds influence the predicted nebular line ratios in photoionization models, and we compare our results with observed line ratios from a sample of [WR] PNe (NGC~5315, NGC~40, BD+30°3639, NGC~6905, and NGC~2867). 

In Section \ref{sec:sec2} we describe our models (blackbody, plane-parallel, and [WR]) and present our differential analysis of their ionizing fluxes. The nebular analysis in which we investigate these models' performance in photoionization models and how they compare with the observations is presented later in Section \ref{sec:sec3}. In Section \ref{sec:limitations}, we discuss an overview on the limitations of early atmosphere models in photoionization studies in the literature, and state the advantages of the approach proposed in this manuscript. In Section \ref{sec:conclusions}, we present our main findings in a summary. 

\section{The ionizing fluxes of CSPNe: the effect of the presence of a stellar wind}
\label{sec:sec2}

In this section, we analyze the emergent ionizing radiation predicted by different CSPNe models. Specifically, we compare the commonly used blackbody and plane-parallel NLTE models with NLTE expanding atmosphere models.  

[WR] stars are complex objects. Their spectra are formed within dense outflows, with velocities ranging from a few hundred to several thousand~km~s$^{-1}$. Consequently, the assumption of a plane-parallel geometry becomes inadequate, and the LTE approximation no longer holds. Computing realistic models for such stars therefore requires significant computational resources. For this reason, expanding atmosphere models are seldom employed as input in photoionization studies.

\subsection{The stellar atmosphere models}
\label{sec:grid_models} 

To construct the blackbody (bb) models\footnote{The bb models used here are also plane-parallel (pp) models; however, we reserve the latter term for those that include detailed radiative transfer calculations (e.g., NLTE).} used in this work, we rely on the fundamental principles of the classical Planck law, which describes the spectral radiance---that is, the intensity of radiation emitted by a blackbody at a given wavelength and temperature. Mathematically, it is expressed as:

\begin{equation}
   B_{\lambda}(T) = \frac{2hc^{2}/\lambda^{5}}{e^{hc/(\lambda kT)} - 1}
\end{equation}

where $B_{\lambda}(T)$ represents the spectral radiance as a function of wavelength ($\lambda$) and temperature ($T$), $h$ is Planck’s constant (6.63$\times$10$^{-27}$~erg~s), $c$ is the speed of light (3$\times$10$^{10}$~cm~s$^{-1}$), and $k$ is Boltzmann’s constant (1.38$\times$10$^{-16}$~erg~K$^{-1}$). The units of $B_{\lambda}(T)$ are erg~cm$^{-2}$~s$^{-1}$~ster$^{-1}$~\AA$^{-1}$. This equation describes the distribution of radiation emitted by a blackbody across different wavelengths, with the peak intensity shifting toward shorter wavelengths as the temperature increases. We then obtained the flux (erg~cm$^{-2}$~s$^{-1}$~\AA$^{-1}$) and luminosity (erg~s$^{-1}$~\AA$^{-1}$) by computing:

\begin{equation}
   F_{\lambda} = B_{\lambda}(T) \int_{0}^{2\pi} d\phi \int_{0}^{\pi/2} \cos\theta \sin\theta\, d\theta = \pi B_{\lambda}(T),
\end{equation}

\begin{equation}
   L_{\lambda} = 4\pi R_{\ast}^{2} F_{\lambda},
\end{equation}

where $R_{\ast}$ is the stellar radius. By varying the temperature parameter in Planck’s law, we generate theoretical spectra that allow us to compare and analyze the ionizing fluxes within this approximation.

For the plane-parallel (pp) approximation, we adopted models from the grid computed by \cite{rauch2003grid}. Their grid provides a comprehensive set of NLTE plane-parallel models covering a wide range of temperatures, luminosities, and surface gravities appropriate for planetary nebulae with hot central stars. In particular, we selected models constructed with “typical” PG~1159 abundance ratios, in which the atmosphere is assumed to consist only of He and CNO elements, therefore consistently representing hydrogen-deficient CSPNe. The grid spans effective temperatures between $T_{\mathrm{eff}} = 40$--190~kK and surface gravities of $\log g = 5$--9. It is worth noting that for models with $T_{\mathrm{eff}} > 100$~kK, the available $\log g$ values range from 6 to 9, whereas for cooler models they start at $\log g = 5$. 

To investigate the influence of stellar winds on the emergent radiation of [WR] stars, we employed the expanding atmosphere models of \cite{keller2011new}.

These models were computed using CMFGEN \citep{hillier1998treatment}, a well-established non-LTE stellar atmosphere code. Within a spherically symmetric expanding outflow, CMFGEN simultaneously solves the radiative transfer equation together with the statistical and radiative equilibrium equations. The code also accounts for line blanketing, an important effect that influences the atmospheric structure of hot stars such as Wolf–Rayet (WR) stars.

The grid from \citet{keller2011new} is well suited for analyzing high-resolution spectra of evolved, hydrogen-deficient post-AGB stars with effective temperatures above 50,000 K. The model parameters include the stellar temperature ($T_{\ast}$), luminosity ($L$), surface gravity ($\log g$), stellar radius ($R_{\ast}$), mass-loss rate ($\dot{M}$), terminal velocity ($v_{\infty}$), and the transformed radius ($R_{t}$), which provides a measure of the wind density and is defined as:

\begin{equation}
\label{eq:transformed_radius}
R_{t} = R_{\ast} \left( \frac{v_{\infty} / 2500,\mathrm{km,s^{-1}}}{\dot{M} / 10^{-4},M_{\odot},\mathrm{yr}^{-1}} \right)^{2/3},
\end{equation}

where a smaller value of $R_{t}$ indicates a denser stellar wind.

In this study, we selected models from two groups, A and B. Group A models represent central stars with masses of 0.5~$M_{\odot}$, covering effective temperatures from 50,000~K to 125,000~K and $\log g$ values from 4.4 to 6.6. Group B models correspond to 0.6~$M_{\odot}$ central stars, with temperatures ranging from 50,000~K to 200,000~K and $\log g$ between 4.0 and 7.0. The mass-loss rate varies between $\log \dot{M} = -6.0$ and $-7.0$ for both groups, depending on the specific model.

Table~\ref{tab:wrmodels} lists the subset of models used in our analysis. They were chosen to span a representative range of temperature, luminosity, and surface gravity, allowing us to investigate how the stellar wind influences the emergent radiation field and how this effect depends on these parameters.

It is worth noting that the \citet{keller2011new} grid does not include models with $T_{\ast} < 80~\mathrm{kK}$ at $\log g \ge 5$, while the \citet{rauch2003grid} models do not extend to $\log g \le 5$. Because cooler CSPN atmosphere models are essential for exploring wind effects over the full temperature range reported in the literature, we supplemented our sample with a single Group A model at $T_{\ast}=50~\mathrm{kK}$, choosing the available $\log g$ value closest to the lower limit of the \citet{rauch2003grid} grid (see Table~\ref{tab:wrmodels}). This introduces a minor $\log g$ mismatch between the two grids, which should be considered when comparing their emergent fluxes.

\begin{table*}
\centering
\caption[Selected {[}WR{]} models from Keller et al. (2011).]{Selected {[}WR{]} models from \cite{keller2011new}.}
\label{tab:wrmodels}
\begin{tabular}{lccccccc}
\hline
\textbf{Modelo} & \textbf{T$_{*}$ {[}10$^3$K{]}} & \textbf{$\log$ (g)} & \textbf{R (R$_{\odot}$)} & \textbf{L(10$^3$L$_{\odot}$)} & \textbf{$\log$  ($\dot{M}$)}{[}M$_{\odot}$ year$^{-1}${]}) & \textbf{v$_{\infty}$(km s$^{-1}$)} & \textbf{R$_t$(R$_{\odot}$)} \\
\hline
 A50.M63.V1000 & 50 & 4.4 & 0.75 & 3.10 & -6.3 & 1000 & 13.80 \\
A50.M73.V1000 & 50 & 4.4 & 0.75 & 3.10 & -7.3 & 1000 & 64.04  \\
 A80.M63.V1000 & 80 & 5.3 & 0.27 & 2.50 & -6.3 & 1000 & 4.90 \\
A80.M73.V1000 & 80 & 5.3 & 0.27 & 2.50 & -7.3 & 1000 & 22.80  \\
A100.M67.V2500 & 100 & 6.0 & 0.12 & 1.20 & -6.7 & 2500 & 7.40 \\
A100.M73.V2500 & 100 & 6.0 & 0.12 & 1.20 & -7.3 & 2500 & 18.70 \\
B150.M65.V2000 & 150 & 6.0 & 0.12 & 7.0 & -6.5 & 2000 & 3.90 \\
B150.M70.V2000 & 150 & 6.0 & 0.12 & 7.0 & -7.0 & 2000 & 10.70 \\
B165.M67.V2000 & 165 & 6.3 & 0.09 & 5.8 & -6.7 & 2000 & 8.10 \\
B165.M70.V2000 & 165 & 6.3 & 0.09 & 5.8 & -7.0 & 2000 & 7.10 \\
\hline
\end{tabular} \\
\flushleft
\footnotesize{The letters \textbf{A} and \textbf{B} correspond respectively to central stars models with masses of 0.5 M$_{\odot}$ and 0.6 M$_{\odot}$. Furthermore, the \textbf{M} represents the mass loss rate and \textbf{V} the terminal velocity adopted for the models. The second column presents the temperature of the models, followed by the surface gravity, radius of the object, luminosity, mass-loss rate, terminal velocity, and transformed radius.}
\end{table*}

In the following section, we present the first part of our analysis, which consists of a qualitative investigation of the influence of stellar winds on the emergent radiation field of CSPNe, and how this effect varies across different temperature ranges. In the second part, we calculate the number of ionizing photons produced by each model and provide a quantitative discussion of these results, together with a broader interpretation of our findings up to this stage. Subsequently, in Section \ref{sec:sec3}, we analyze a sample of [WR] stars using observed emission-line ratios.

\subsection{The differential analysis of the ionizing spectra}

\citet{rauch2003grid} demonstrated substantial differences between plane-parallel and simple blackbody models by comparing their emergent fluxes. In the ultraviolet region of the spectrum, the fluxes diverge as photon energy increases. In the EUV regime, more than one blackbody model would be required to reproduce the flux of a single plane-parallel model. However, those analyses were performed for H-rich CSPNe, where the stellar wind plays a negligible role. In this section, we aim to extend that scenario to [WR] stars, in which the stellar wind has a major influence, as their spectra are formed within an expanding atmosphere. To this end, we include expanding atmosphere models as an additional step beyond Rauch’s analysis, allowing us to account for the effects of the wind on the emergent radiation field.  

Our initial approach involves a qualitative examination of the ionizing flux produced by each stellar atmosphere model. For every [WR] model selected from the grid of \citet{keller2011new}, we chose a corresponding plane-parallel model from \citet{rauch2003grid} and a blackbody model with the same temperature, luminosity, and surface gravity. Additionally, for each group of models, we included a second [WR] model with a lower mass-loss rate to investigate how this parameter affects the ionizing flux.  

Figures~1–3 illustrate that the EUV spectra of [WR] models differ significantly from their plane-parallel and blackbody counterparts (note that only the continua from \citealt{keller2011new} models are shown below 900~\AA). The mass-loss rate exerts a strong influence on the EUV emission: higher mass-loss rates (i.e., denser winds) alter the number of high-energy photons escaping from the star. This behavior is explored quantitatively in the next section, where we calculate the number of ionizing photons reaching the nebula, expressed as $\log Q$ values for key ionization edges (H, He~I, and He~II).

\begin{figure*}
	\includegraphics[width=\textwidth]{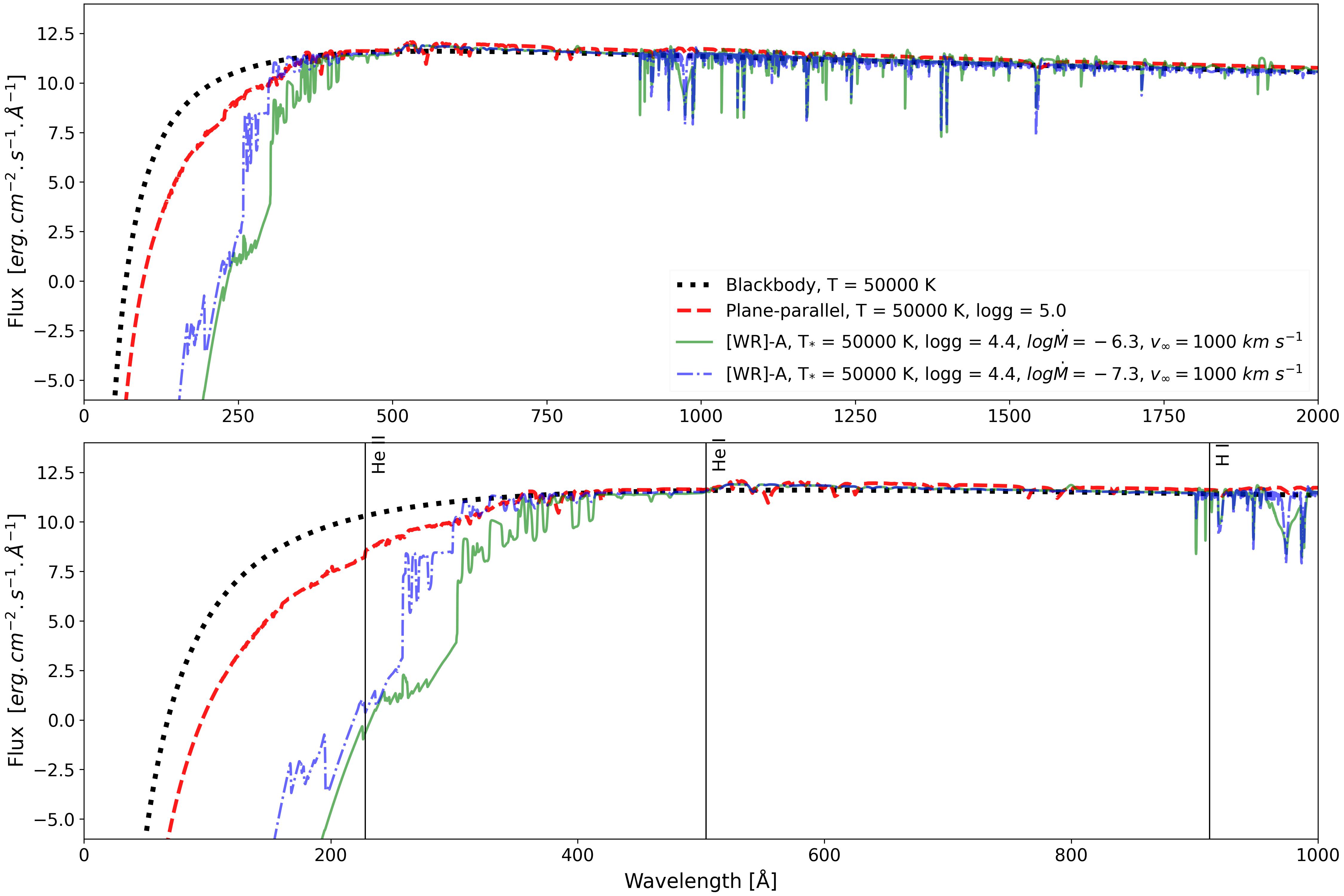}
    \caption{Comparison of the theoretical spectra in the ultraviolet region. For {[}WR{]} models, the mass-loss rate is displayed. The vertical axis represents the flux on a logarithmic scale, while the horizontal axis displays the wavelengths in angstroms. In the lower panel, we highlight the EUV region. Solid vertical lines indicate the wavelengths that correspond to the photoionization of H I (912 \AA), He I (504 \AA), and He II (228 \AA). Note the differences between the models in the EUV despite having the same temperature, luminosity, and log g.
    \textbf{[WR]-A refers to the [WR] models labeled "A" in \cite{keller2011new}.}}
    \label{fig:fluxA50}
\end{figure*}

In Figure~\ref{fig:fluxA50}, we present [WR] models representing central stars with masses of $0.5~M_{\odot}$ (model A) and an effective temperature of 50~kK, alongside their corresponding blackbody and plane-parallel approximations. The primary discrepancies between these models become evident in the EUV region, particularly between the He~I ($504$~\AA) and He~II ($228$~\AA) ionization thresholds. The [WR] model with the lower mass-loss rate ($\log \dot{M} = -7.3$) exhibits a higher EUV continuum, suggesting that a weaker stellar wind causes the model atmosphere to approach the plane-parallel limit. In contrast, the blackbody model predicts a flux that extends substantially beyond the He~II ionization threshold when compared to the other models.  

The model with the stronger stellar wind ($\log \dot{M} = -6.3$) displays pronounced attenuation of its ionizing continuum already between $504$~\AA\ and $228$~\AA, indicating that a denser wind more effectively suppresses EUV radiation. This finding reinforces the conclusions of \citet{rauch2003grid}, while further highlighting that in [WR] atmospheres, the presence of a stellar wind must be taken into account, as it reshapes the emergent spectral energy distribution by reducing the EUV flux.

\begin{figure*}
	\includegraphics[width=\textwidth]{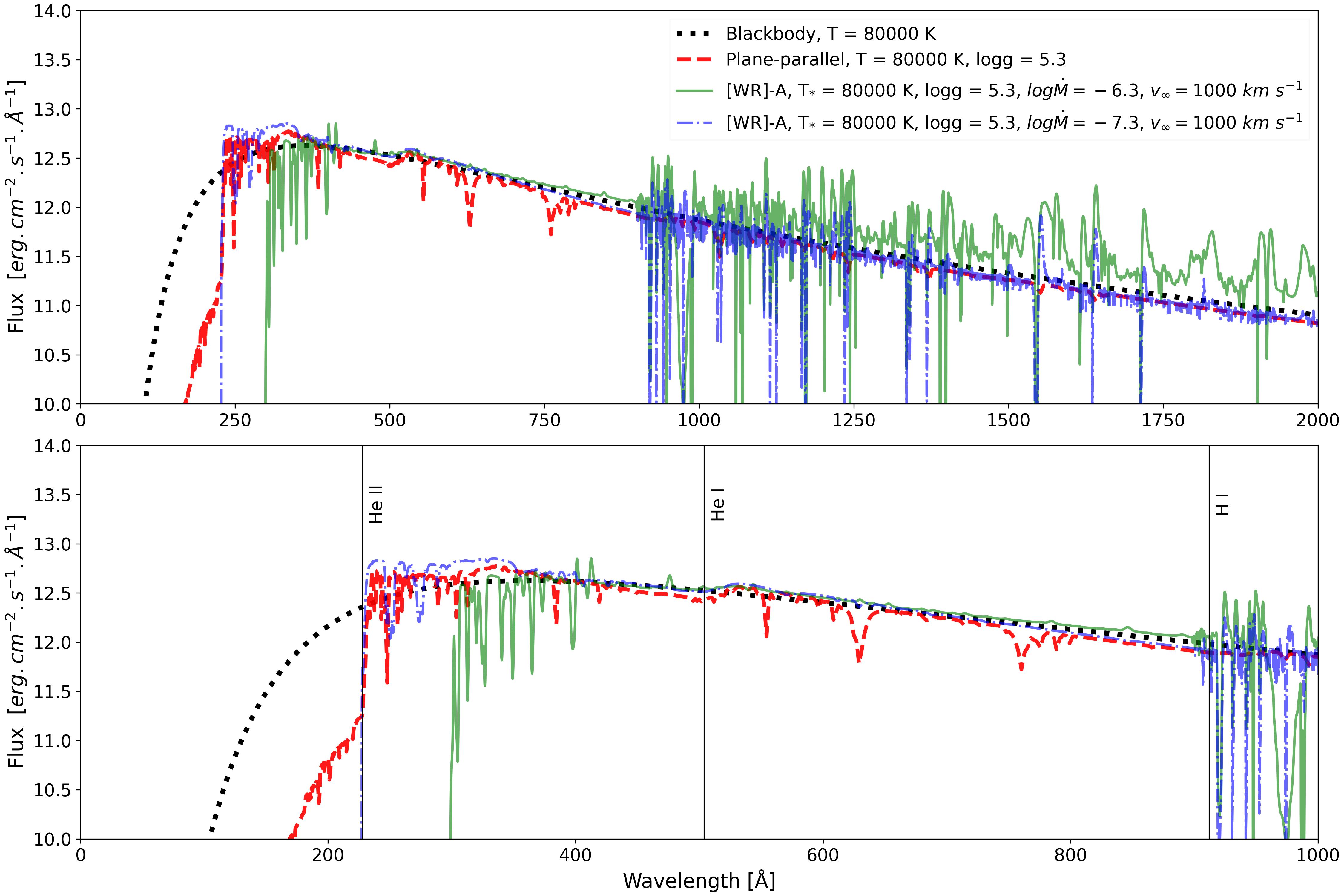}
    \caption{Same as Fig. \ref{fig:fluxA50}, but for T$_{*}$ = 80000 K, log g = 5.3 and v$_{\infty}$ = 1000 km $s^{-1}$. \textbf{[WR]-A refers to the [WR] models labelled "A" in \cite{keller2011new}.}}
    \label{fig:fluxA80}
\end{figure*}

In Figure~\ref{fig:fluxA80}, we show the theoretical fluxes for the [WR] model with an effective temperature of 80~kK, alongside its corresponding blackbody and plane-parallel counterparts. In this case, the discrepancies among the models become pronounced in the EUV region, particularly between 504~\AA\ and 228~\AA. The [WR] model with the lower mass-loss rate ($\log \dot{M} = -7.3$) presents a flux distribution very similar to the plane-parallel model, once again indicating that a weaker wind drives the atmosphere toward the plane-parallel regime. In contrast, the blackbody flux extends well beyond the He~II ionization threshold compared to the other models.  

The [WR] model with the higher mass-loss rate ($\log \dot{M} = -6.3$) exhibits a strong attenuation in its ionizing continuum, beginning earlier than in the other cases (between 504~\AA\ and 228~\AA). This result not only reinforces the conclusions of \citet{rauch2003grid}, but also highlights that in [WR] stars, the stellar wind must be considered, as it shapes the radiation field emerging from the central star by attenuating the far-UV intensity. It is also noticeable that the [WR] models display a slightly higher flux level at redder wavelengths. This is a consequence of flux redistribution: since all models have the same luminosity and the [WR] models experience EUV flux suppression due to the stellar wind, the lost energy is redistributed toward longer wavelengths, enhancing the continuum in that region.

\begin{figure*}
	\includegraphics[width=\textwidth]{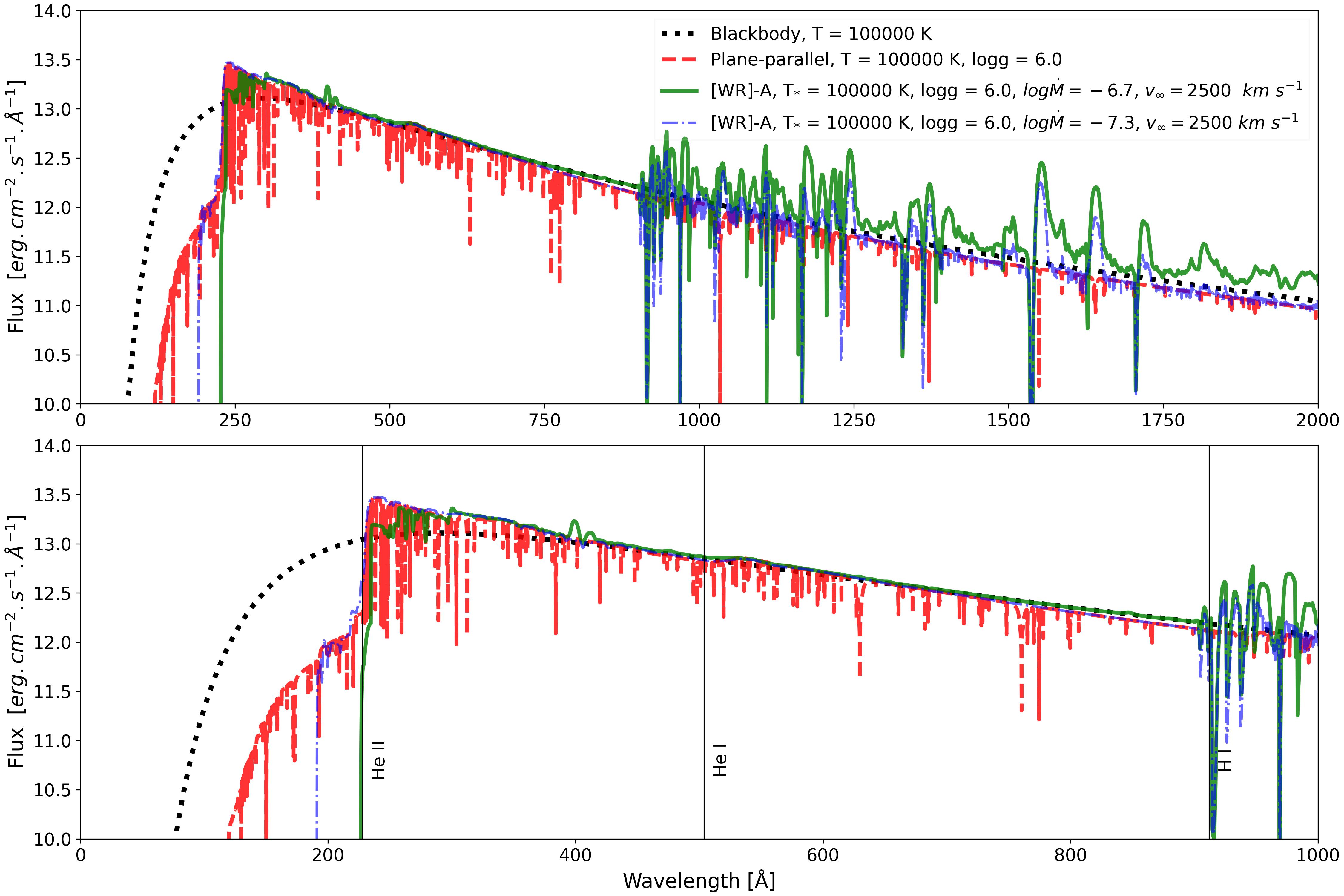}
    \caption{Same as Fig. \ref{fig:fluxA50}, but for T$_{*}$ = 100000 K, log g = 6.0 and v$_{\infty}$ = 2500 km $s^{-1}$. \textbf{[WR]-A refers to the [WR] models labelled "A" in \cite{keller2011new}.}}
    \label{fig:fluxA100}
\end{figure*}

In Figure~\ref{fig:fluxA100}, we compare the third set of models in our sample, each with an effective temperature of 100~kK. The overall behavior of the theoretical flux follows the same pattern observed for the 80~kK case, with clear distinctions between the blackbody, plane-parallel, and [WR] models. The mass-loss rate again plays a crucial role in shaping the emergent radiation, as higher wind densities lead to stronger attenuation of the flux. The blackbody model extends farther into the EUV region than the others, while the plane-parallel model more closely resembles the [WR] model with the lowest mass-loss rate.  

In the previous temperature case, the [WR] model with the highest wind density showed a pronounced drop near 300~\AA, just before the He~II ionization threshold, followed by the second [WR] model with a weaker wind. In the hotter models presented here, however, this attenuation occurs beyond the He~II ionization threshold, reflecting the increased transparency of the atmosphere at higher temperatures.

\begin{figure*}
	\includegraphics[width=\textwidth]{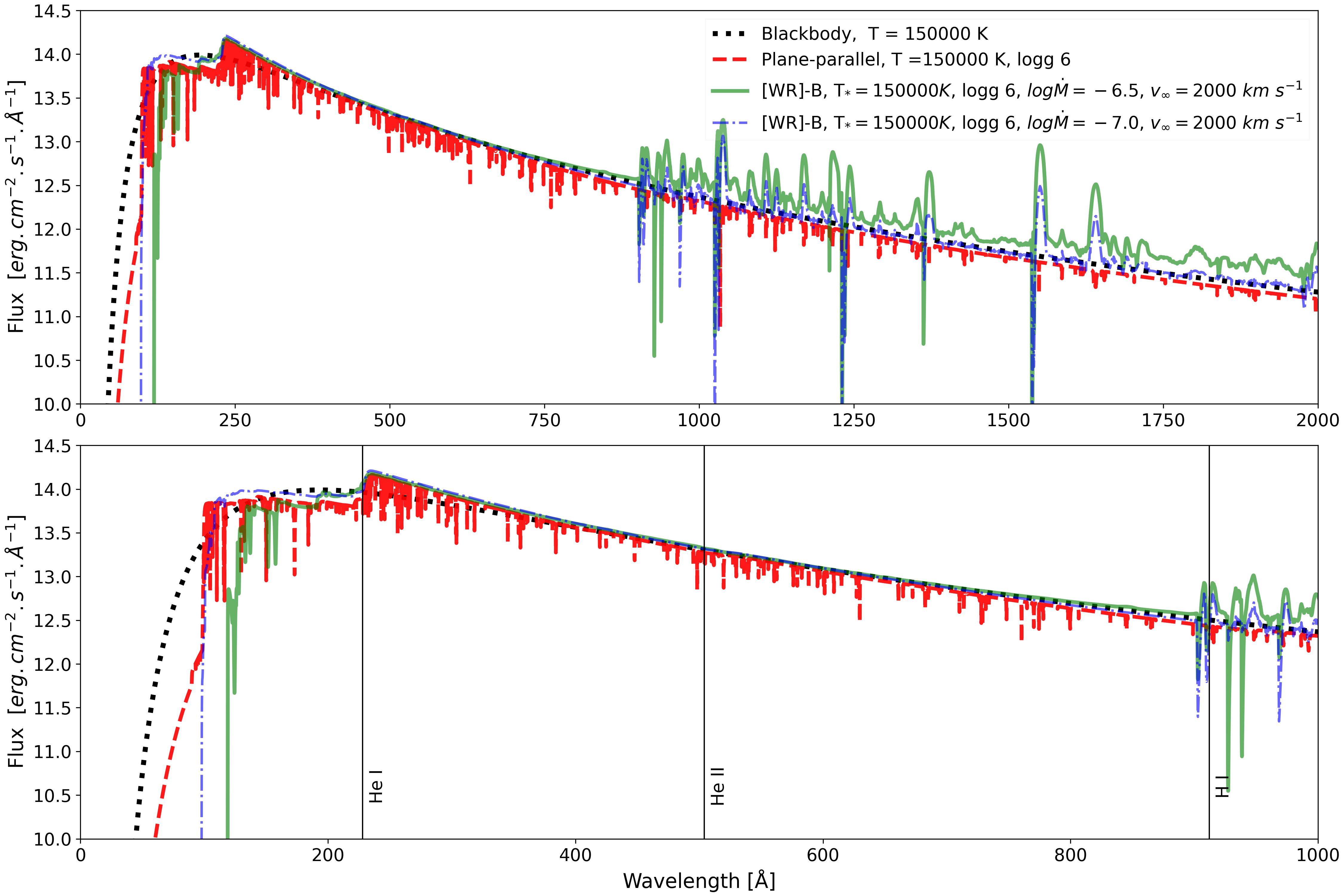}
    \caption{Same as Fig. \ref{fig:fluxA50} and \ref{fig:fluxA100}, but for T$_{*}$ = 150000 K, log g = 6.0 and v$_{\infty}$ = 2000 km $s^{-1}$.}
    \label{fig:fluxB150}
\end{figure*}

In Figure~\ref{fig:fluxB150}, we compare the fourth set of models in our sample, each with an effective temperature of 150~kK, together with their corresponding blackbody and plane-parallel approximations. For this temperature range, [WR] models differing by exactly one dex in mass-loss rate were not available, so we adopted the largest difference possible within the grid. The models exhibit comparable flux levels between the H~I (912~\AA) and He~II (228~\AA) ionization thresholds. The impact of the mass-loss rate becomes noticeable only below 228~\AA, where it introduces a modest attenuation of the ionizing flux.

\begin{figure*}
	\includegraphics[width=\textwidth]{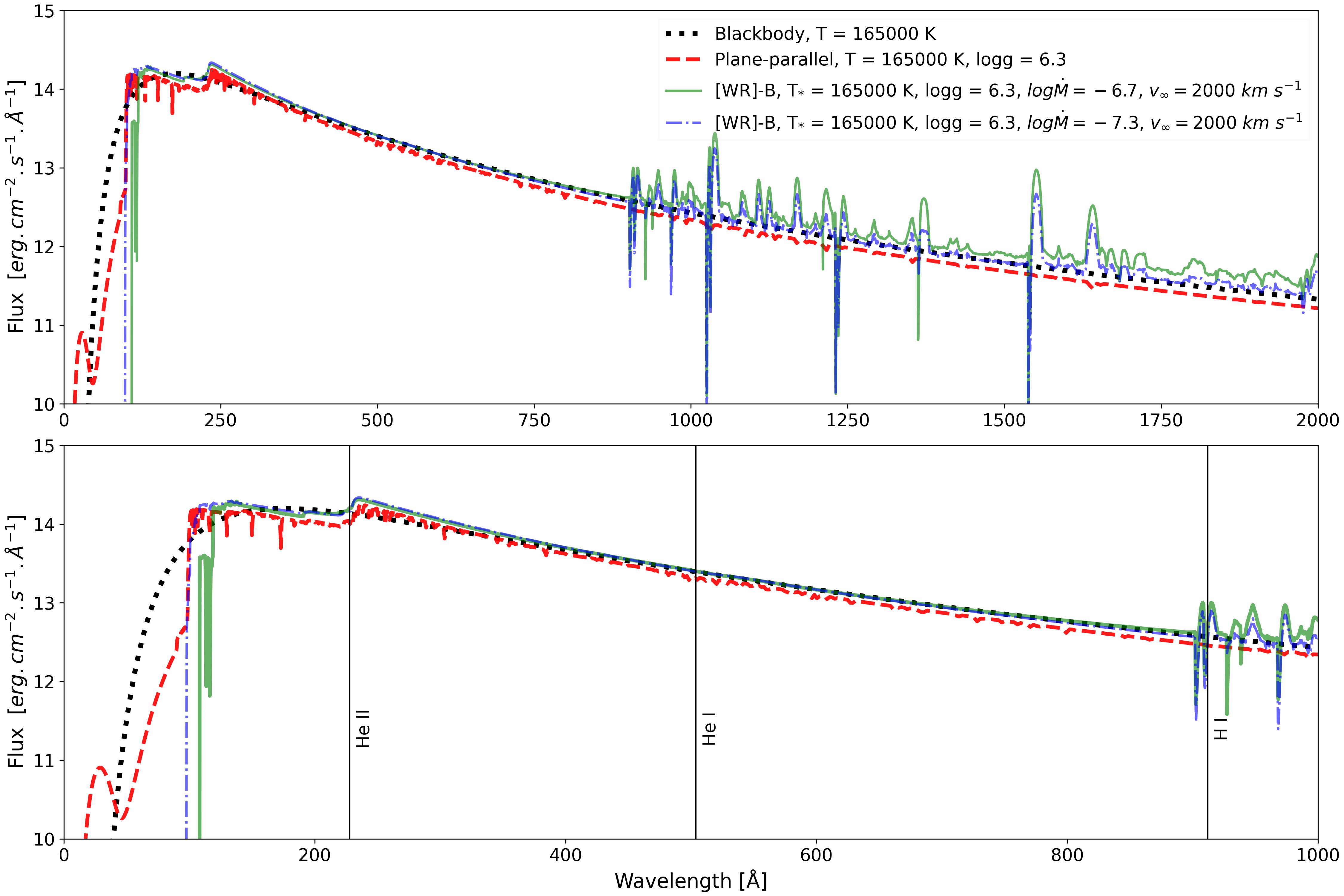}
    \caption{Same as Fig. \ref{fig:fluxA50} and \ref{fig:fluxA100}, but for T$_{*}$ = 165000 K, log g = 6.3 and v$_{\infty}$ = 2000 km $s^{-1}$.}
    \label{fig:fluxB165}
\end{figure*}

In Figure~\ref{fig:fluxB165}, we present the comparison for the final set of models in our sample, each with an effective temperature of 165~kK. The same general trends persist: the blackbody model extends farther into the EUV region, while the model with a higher mass-loss rate exhibits a reduced EUV flux. Conversely, the [WR] model with the lower mass-loss rate closely approaches the plane-parallel model in terms of flux distribution.  

Overall, our results demonstrate that the presence of a stellar wind plays a crucial role in shaping the emergent EUV flux, acting as a significant source of opacity for ionizing photons. This effect is most pronounced at temperatures below roughly 100~kK. Consequently, analyzing PNe with [WR] central stars, particularly the cooler [WCL] types, using plane-parallel or blackbody-based ionizing sources is not only inconsistent but may lead to systematic errors.  

To further quantify the influence of stellar wind strength and temperature on the emergent EUV flux, the next section presents a detailed analysis based on the number of ionizing photons, expressed as $\log Q$.

\subsection{The number of ionizing photons}

A quantitative method to evaluate the amount of ionizing radiation emitted by the central star toward the nebula is through the calculation of the number of ionizing photons per second ($Q$). This parameter can be directly incorporated into photoionization models of PNe and is determined using the following expression:

\begin{equation}
    Q = \int_{\nu_{0}}^{\infty} \frac{L_{\nu}}{h\nu} \, d\nu
    \label{eq:number_of_ioniphot}
\end{equation}

In the above equation, $h$ is Planck’s constant, $\nu_{0}$ represents the ionization threshold frequency for the atom or ion, and $L_{\nu}$ is the monochromatic luminosity.  

Table~\ref{tab:numberofIPHO} lists the number of ionizing photons for H~I, He~I, and He~II for all stellar atmosphere models used in our analysis.

\begin{table}
\centering
\caption[Number of ionizing photons]{Number of ionizing photons}
\label{tab:numberofIPHO}
\begin{tabular}{c|ccccccc}
\hline
\textbf{Models} & \textbf{$\log$ Q$_{0}$($\lambda$912)} & \textbf{$\log$ Q$_{1}$($\lambda$504)} & \textbf{$\log$ Q$_{2}$($\lambda$228)} \\
\hline
A50.M63.V1000 & 47.40 & 46.32 & 32.87 \\
A50.M73.V1000 & 47.41 & 46.49 & 34.28 \\
blackbody & 47.40 & 46.70 & 44.30 \\
plane-parallel & 47.32 & 46.40 & 41.76 \\
\hline
A80.M63.V1000 & 47.29 & 46.81 & 33.57 \\
A80.M73.V1000 & 47.36 & 47.02 & 43.15 \\
blackbody & 47.37 & 47.03 & 45.71 \\
plane-parallel & 47.27 & 46.93 & 44.27 \\
\hline
A100.M67.V2500 & 47.01 & 46.76 & 42.56 \\
A100.M73.V2500 & 47.02 & 46.79 & 44.67 \\
blackbody & 47.02 & 46.78 & 45.82 \\
plane-parallel & 46.98 & 46.75 & 44.72 \\
\hline
B150.M65.V2000 & 47.61 & 47.48 & 46.75 \\
B150.M70.V2000 & 47.64 & 47.53 & 46.90 \\
blackbody & 47.61 & 47.50 & 46.97 \\
plane-parallel & 47.58 & 47.46 & 46.80 \\
\hline
B165.M67.V2000 & 47.75 & 47.65 & 47.09 \\
B165.M70.V2000 & 47.77 & 47.67 & 47.14 \\
blackbody & 46.86 & 47.20 & 46.90 \\
plane-parallel & 47.49 & 47.39 & 46.85 \\
\hline
\end{tabular}\\
\flushleft
\footnotesize{In this table, the letters \textbf{A} and \textbf{B} account for {[}WR{]} models with masses respectively of 0.5 M$_{\odot}$ and 0.6 M$_{\odot}$, while \textbf{M}  represents the logarithm of the mass-loss rate, \textbf{V} the terminal velocity of the wind, and $\log$ Q$_{0}$, $\log$ Q$_{1}$, and $\log$ Q$_{2}$ are the logarithm of the number of ionizing photons of H I (912 \AA), He I (504 \AA), and He II (228 \AA) respectively.}
\end{table}

From Table~\ref{tab:numberofIPHO}, it is evident that the stellar wind acts as a significant source of opacity for the radiation field emerging from [WR] central-star models with $T_{\ast} \leq 80$~kK. In these cases, the model mass-loss rate produces a substantial reduction, up to $\sim10$~dex, in the number of ionizing photons emitted in the highest-energy portion of the EUV ($\lambda \leq 228$~\AA), i.e., for photons energetic enough to ionize He~\textsc{ii} in the nebula. The wind also contributes to a decrease in the number of He~\textsc{i}-ionizing photons, whereas the H~\textsc{i}-ionizing photon counts from the [WR] and plane-parallel models are comparable (and in some cases slightly higher for the [WR] models).  

When the temperature increases to 100~kK, the numbers of ionizing photons for H~\textsc{i} and He~\textsc{i} produced by the different models become roughly similar, implying that, in this respect, any of the models would affect photoionization calculations in a comparable manner. The same pattern holds for the models at 150 and 165~kK. Nevertheless, the stellar wind continues to affect the EUV photon budget ($\lambda \leq 228$~\AA) even at 165~kK, attenuating the He~\textsc{ii}-ionizing flux relative to the plane-parallel and blackbody cases. This can lead to overpredictions of certain nebular line intensities, particularly when using a blackbody approximation,although the discrepancy is far less severe than at lower temperatures.  

In the next section, we employ photoionization models to reproduce observed line intensities (I$_{\lambda}$/I$_{H\beta}$) for a sample of PNe hosting [WR] central stars, and we assess whether [WR] atmosphere models provide a better ionizing source than blackbody or plane-parallel approximations.

\section{IONIZING RADIATION AND NEBULAR ANALYSIS FOR A SAMPLE OF PNe}
\label{sec:sec3}

In this section, we employ the ionizing fluxes produced by different stellar atmosphere models as input for photoionization simulations. We predict nebular emission-line ratios for several ions and compare them with observed data for five well-known PNe: NGC~40 ([WC8]), NGC~5315 ([WO4]), BD+30°3639 ([WC9]), NGC~6905 ([WO2]), and NGC~2867 ([WO2]). The models were computed using the \textsc{cloudy} code \citep{ferland2017}, version~07.02.01. \textsc{Cloudy} solves the ionization balance, statistical equilibrium, and thermal balance of both gas and dust in the nebula, coupled with the radiative transfer equations. The simultaneous solution of these non-linear coupled equations yields a self-consistent structure for the nebula based on user-defined input parameters.  

For each model, we specified the total hydrogen density ($n_{\mathrm{H}}$), the nebular inner radius ($r_{0}$), and the distance to Earth. All models were assumed to be spherically symmetric. To represent the chemical composition, we used the \texttt{abundance planetary nebulae} command in \textsc{cloudy}, which adopts a composition typical of planetary nebulae \citep{ferland2017}. Although \textsc{cloudy} allows the inclusion of numerous detailed physical features, constructing a fully optimized model for each object would demand substantial computational effort and lies beyond the scope of this study. Instead, our objective is to explore, for the first time, the impact of [WR] stellar winds on nebular conditions and emission-line ratios, relative to simpler ionizing sources, and to determine which approach best reproduces the observed data in a first approximation.  

For each configuration, we obtained the predicted nebular line ratios (I$_{\lambda}$/I$_{H\beta}$) and compared them with extinction-corrected observed values normalized to H$_{\beta}$. To quantify the agreement between model predictions and observations, we computed the root mean square (rms) deviation using the following expression:

\begin{equation}
    rms = \sqrt{\frac{1}{N}\left(1 - \frac{\text{model}}{\text{observed}}\right)^{2}}
\end{equation}

where $N$ denotes the number of emission lines analyzed for each PN in our sample, “model” refers to the predicted line ratio from the photoionization calculation, and “observed” corresponds to the measured value from the literature.

Initially, all parameters used to construct the photoionization models were kept fixed. The total hydrogen density was first assumed to decrease with the square of the distance from the central star, as preliminary tests indicated that this configuration provided slightly better agreement with the observed data. In the subsequent test iterations, we allowed the hydrogen density to vary in order to improve the match between the modeled and observed line ratios. Once the best-fitting model was obtained, the optimized hydrogen density was fixed, and we proceeded to adjust the nebular radius. The initial values of the nebular radii were adopted from estimates available in the literature, while the initial assumption for the total hydrogen density (log~$n_{\mathrm{H}}$) was set to 4.0, which is a representative value for planetary nebulae \citep{osterbrock2006astrophysics}.  

The following section provides a detailed discussion of each object in our sample.

\subsection{NGC 5315}
\label{sec:sec3.1}

The planetary nebula NGC~5315 is characterized as a relatively low-excitation object hosting a [WO4] central star \citep{acker2003quantitative}, with an estimated stellar mass of 0.57~M$_{\odot}$ \citep{marigo2003probing}. The central star exhibits a mass-loss rate of $\log\dot{M} = -6.33$~(M$_{\odot}$~yr$^{-1}$), a terminal velocity $v_{\infty} = 2400$~km~s$^{-1}$, and an effective temperature of 76,000~K \citep{marcolino2007detailed}. Morphologically, NGC~5315 is a compact nebula, with diameter estimates ranging between $4^{\prime\prime}$ and $6^{\prime\prime}$, showing a slightly elliptical geometry and a broken ring-like structure \citep{pottasch2002abundances}.  

To model this nebula, we adopted the stellar parameters derived by \citet{marcolino2007detailed} using the CMFGEN code \citep{hillier1998treatment}, whose detailed NLTE analysis provides a robust reproduction of the UV and optical spectra of the central star. Simpler approaches, such as blackbody or plane-parallel prescriptions, fail to reproduce the characteristic [WR] emission features, making the CMFGEN-derived parameters a more reliable representation of the star in NGC~5315. The adopted stellar parameters are: temperature $T_{*} = 76,000$~K, luminosity $L = 5000$~L$_{\odot}$, radius $R_{*} = 0.40$~R$_{\odot}$, and surface gravity $\log g = 5.0$.  

The spectral energy distributions (SEDs) from the different model grids, CMFGEN, plane-parallel, and blackbody, were originally given in distinct flux units and normalization schemes. To ensure consistency and proper input into the photoionization code, all SEDs were converted to the same physical units of flux (erg~cm$^{-2}$~s$^{-1}$~\AA$^{-1}$) at the stellar surface. The total flux of each model was then scaled according to the stellar parameters adopted for NGC~5315, using the stellar mass of 0.57~M$_{\odot}$ from \citet{marigo2003probing}, ensuring that luminosities remained consistent across all models. For the other planetary nebulae in our sample, fluxes were scaled following the same procedure.  

For comparison, we selected a hydrogen-deficient plane-parallel model from \citet{rauch2003grid} and a blackbody model, both computed with the same temperature, surface gravity, and luminosity as the [WR] star. The adopted distance from Earth for all photoionization simulations of this nebula was 2600~pc \citep{peimbert2004physical}. The results for NGC~5315 are presented and discussed below.

\subsubsection{Results}

\begin{figure*}
	\includegraphics[width=\textwidth]{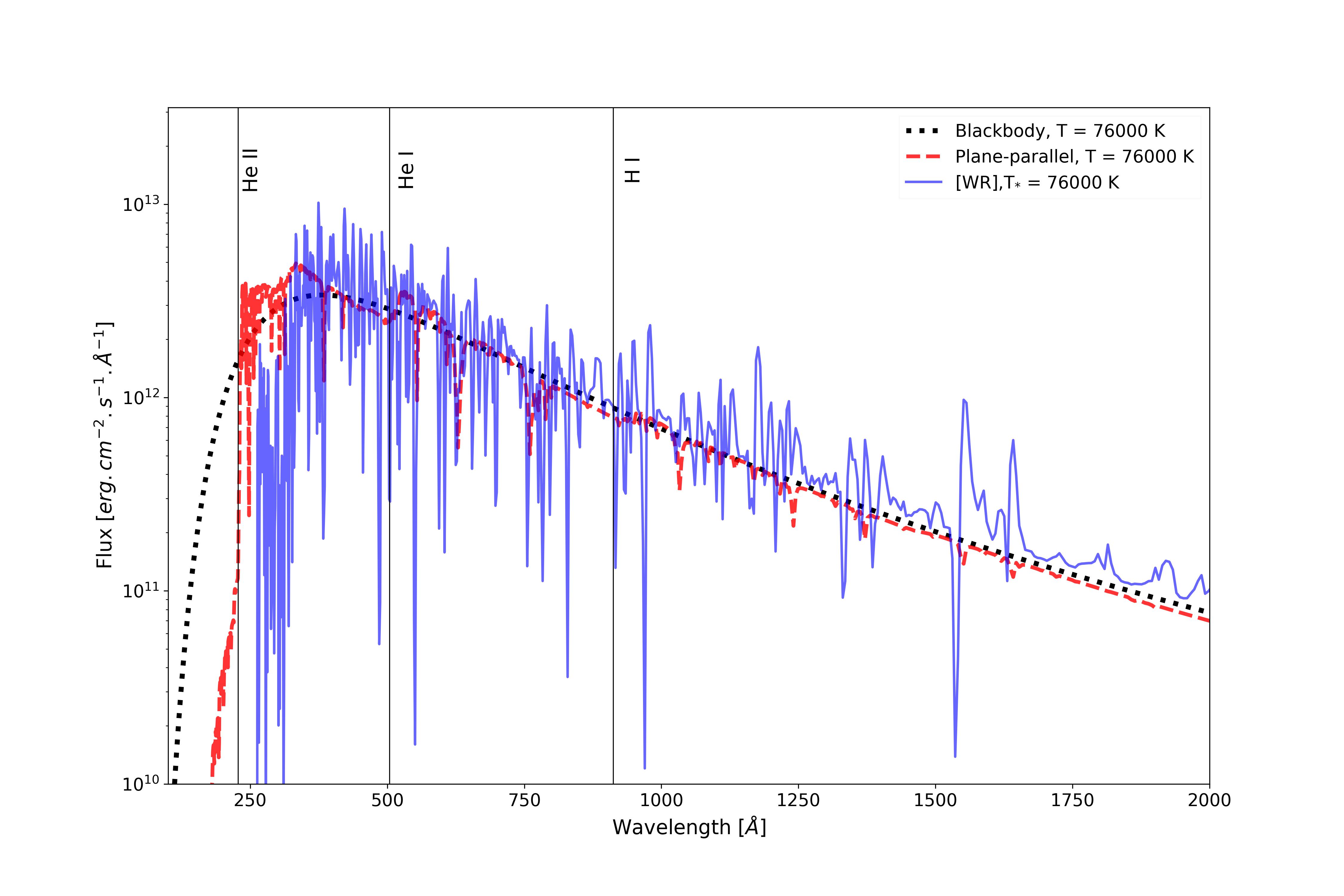}
    \caption{NGC 5315 - Comparison among the theoretical spectra in the  FUV and EUV region. The vertical axis represents the flux on a logarithmic scale, while the horizontal axis displays the wavelengths in angstroms. Solid vertical lines indicate the wavelengths that correspond to the photoionization of HI (912 \AA), He I (504 \AA), and He II (228 \AA).}
    \label{fig:NGC5315fluxio}
\end{figure*}

In Figure~\ref{fig:NGC5315fluxio}, we present a comparison among the theoretical spectra employed in our photoionization models for the PN NGC~5315. It is important to emphasize that all model SEDs represent the intrinsic stellar flux incident on the nebula (i.e., not scaled to the Earth’s distance). As a result, the UV fluxes in the figure appear significantly higher than the observed ones, a condition that applies to all objects in our sample. The scaling to Earth-based fluxes is performed only when comparing the photoionization model outputs with the observational data, using the corresponding distance to each object.  

A visual inspection of Figure~\ref{fig:NGC5315fluxio} reveals that, consistent with the results discussed in previous sections, substantial differences arise among the models for the adopted temperature, becoming more pronounced toward the EUV region. Between 912~\AA\ and 504~\AA, the plane-parallel model produces a weaker ionizing flux compared to both the blackbody and [WR] models. Conversely, below 228~\AA, the [WR] model displays a sharp decline, the plane-parallel model drops near the He~II ionization threshold, while the blackbody flux extends well beyond it. The number of ionizing photons ($\log Q$) for the blackbody model are: H~I = 47.36, He~I = 47.18, and He~II = 45.78. For the plane-parallel model, these values are 47.34, 47.22, and 44.39, and for the [WR] model: 47.38, 47.15, and 38.23. These results reinforce the previous argument that the stellar wind introduces an additional opacity source for the emerging radiation field, as evidenced by the strong reduction in the number of He~II-ionizing photons in the [WR] case.  

The final set of nebular parameters adopted for NGC~5315 was determined through an iterative procedure aimed at reproducing the main observed emission-line ratios. The best-fitting model corresponds to a stellar luminosity of 5000~L$_{\odot}$, a distance of 2600~pc, a nebular inner radius of $\log r_{0} = 17.05$~(cm), and a total hydrogen density of $\log n_{\mathrm{H}} = 4.5$. For consistency, the same $r_{0}$ and $n_{\mathrm{H}}$ values were used for all ionizing sources (blackbody, plane-parallel, and [WR]), ensuring that any variation in the predicted line ratios arises solely from the differences in their ionizing continua. For all nebulae analyzed in this study, the inner radius and hydrogen density were constrained in a similar iterative process.

We used a sample of nebular line ratios measured by \citet{peimbert2004physical}, in which all observed values were corrected for extinction and reddening, and normalized to H$_{\beta} = 100$. The extinction-corrected H$_{\beta}$ flux reported in their work is I$_{H_\beta} = 1.289 \times 10^{-10}$~erg~cm$^{-2}$~s$^{-1}$. Our photoionization models predict I$_{H_\beta}$ values of $6.67\times10^{-11}$~erg~cm$^{-2}$~s$^{-1}$ for the blackbody model, $4.04\times10^{-11}$~erg~cm$^{-2}$~s$^{-1}$ for the plane-parallel model, and $1.12\times10^{-10}$~erg~cm$^{-2}$~s$^{-1}$ for the [WR] model. This comparison indicates a good agreement between the expanding atmosphere model and the observed value.  

\begin{figure*}
	\includegraphics[width=\textwidth]{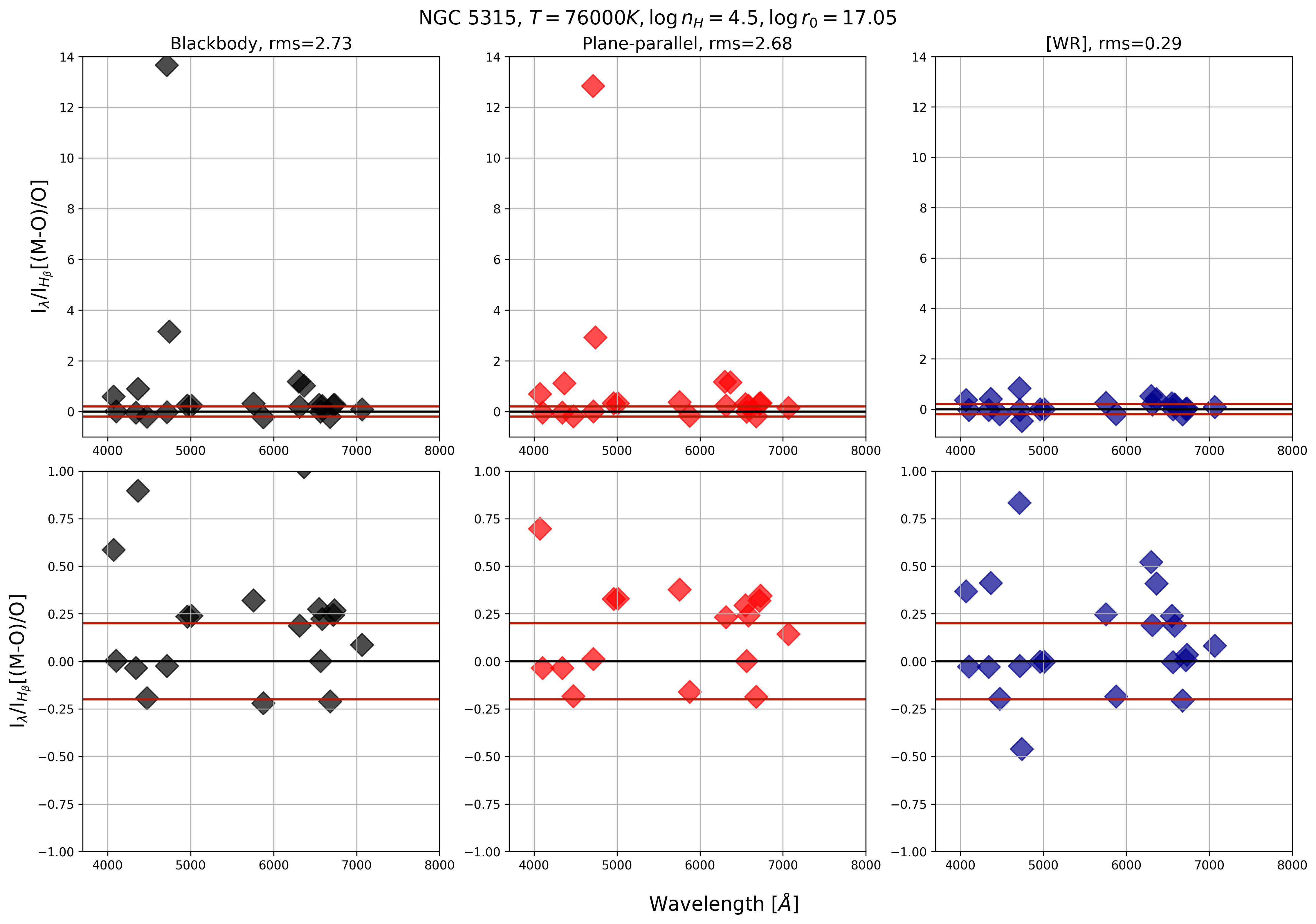}
    \caption{Comparison of model and observed line intensities of the PN NGC 5315 (normalized to H$\beta$). (M) indicates the model-predicted line and (O) the observed line. The points represent the lines displayed in Table \ref{tab:razaolinhasNGC5315}, organized by increasing wavelength in the 4000-8000 \AA interval. The top panel shows all the points while the lower panel zooms in between -100\% and 100\%. Horizontal lines indicating a discrepancy of 20\% are used to guide the eye. Overall, the Cloudy model predictions with the [WR] model as an ionizing source reproduce the observations better visually and through the rms (0.29).}
    \label{fig:NGC5315lineratios}
\end{figure*}

In Figure~\ref{fig:NGC5315lineratios}, we present a comparison between the predicted and observed emission-line intensities for the PN NGC~5315. A visual inspection of the figure reveals that the photoionization model adopting the [WR] stellar atmosphere as the ionizing source provides the best match to the observations, with most lines reproduced within an uncertainty of $\pm$20\%. (For completeness, the numerical results are listed in Appendix~\ref{ap:appendixA}.) The plane-parallel model achieves a slightly better fit than the blackbody model but still fails to reproduce key oxygen features such as [O~III]~$\lambda\lambda$4959,~5007~\AA, which are well matched by the [WR]-based model. Overall, both hydrogen and helium lines are satisfactorily reproduced regardless of the ionizing source, whereas certain forbidden lines, such as [O~II]~$\lambda$7319.5 and 7330.2~\AA, are poorly reproduced in all cases. These lines were therefore treated as outliers, as our focus is to assess the conditions where specific models succeed or fail (see, e.g., \citealt{lee2008photoionization}, for a discussion on the difficulty of reproducing such lines).


For this PN, we conclude that adopting a more realistic stellar atmosphere model leads to a better overall performance of the photoionization model in reproducing the observed nebular line ratios (see Table \ref{tab:razaolinhasNGC5315}). The predictions obtained with the other models are comparatively similar. Therefore, for this object, we infer that the stellar wind plays a crucial role in the nebular photoionization process, not only modifying the emergent flux of the central star but also influencing the nebular emission as a whole.

\subsection{NGC 40}

The planetary nebula (PN) NGC~40, also known as the Bow-Tie Nebula, is a low-excitation object \citep{aller1979spectroscopic}, illuminated by a {[}WC8{]} Wolf-Rayet central star \citep{acker2003quantitative}, with an estimated stellar mass of 0.57~M$_{\odot}$ \citep{monteiro20113d}. 

Distance estimates for this nebula range from 800 to 1400~pc (e.g., \citealt{acker2003quantitative,marcolino2007detailed,monteiro20113d}). For consistency, we adopted the expanding atmosphere model computed by \citet{marcolino2007detailed} in one of our photoionization models for NGC~40, and used the same distance of 1400~pc as assumed in their study.

\subsubsection{Results}

The central star model for this object was computed by \citet{marcolino2007detailed} using the \texttt{CMFGEN} code. The adopted model includes the following key stellar parameters: an effective temperature of 73,000~K ($T_{*}$), a luminosity of 5,000~L$_{\odot}$, a stellar radius of 0.43~R$_{\odot}$, and a surface gravity of $\log g = 5.0$. The stellar wind is characterized by a mass-loss rate of $\log \dot{M} = -6.25$~(M$_{\odot}$~yr$^{-1}$) and a terminal velocity of 1,000~km~s$^{-1}$. For consistency with the flux normalization described in Section~\ref{sec:sec3.1}, we adopted a stellar mass of 0.57~M$_{\odot}$, following \citet{monteiro20113d}.

\begin{figure*}
	\includegraphics[width=\textwidth]{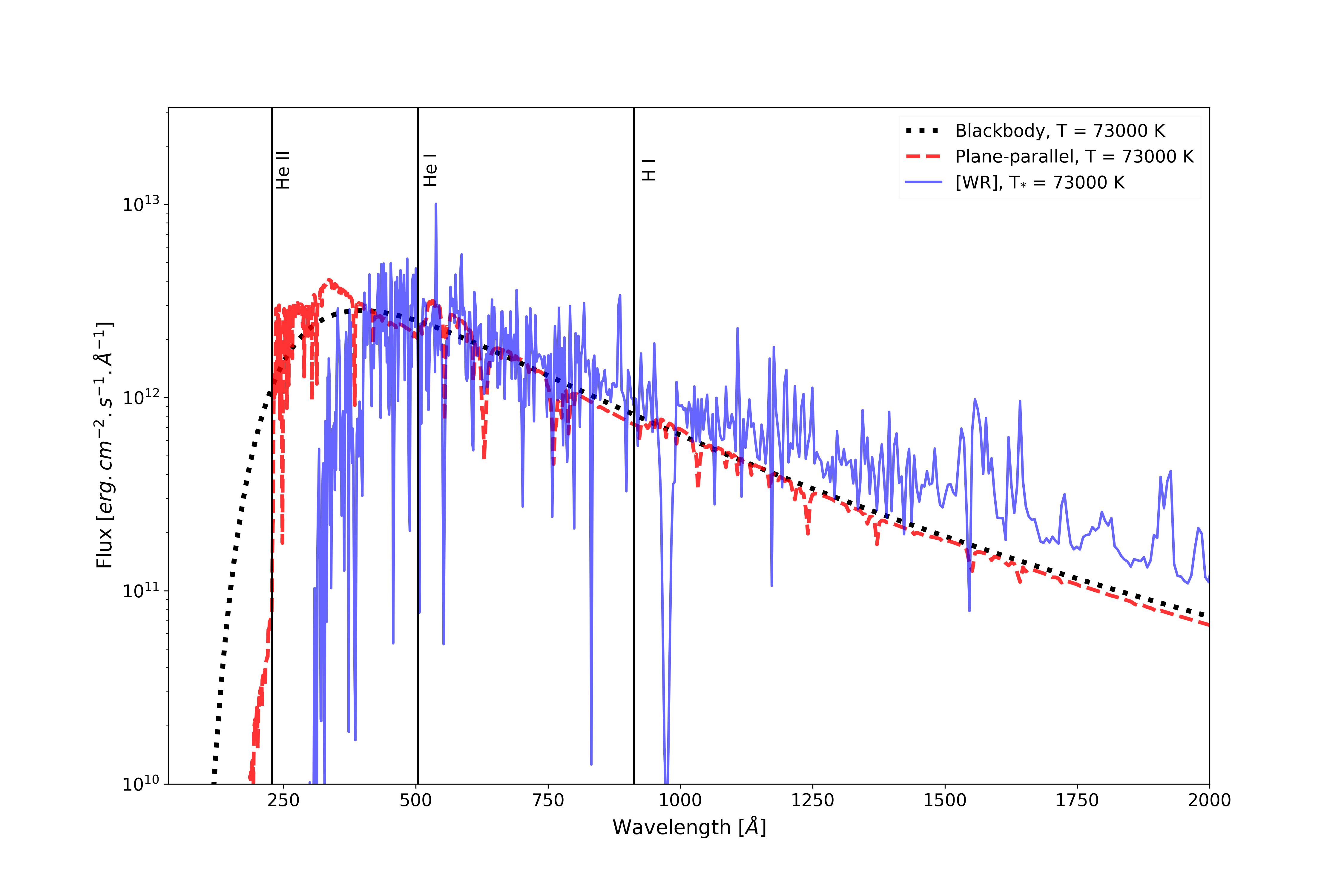}
    \caption{Same as Fig. \ref{fig:NGC5315fluxio}, but for NGC 40.}
    \label{fig:NGC40fluxio}
\end{figure*}

In Figure~\ref{fig:NGC40fluxio}, we present a comparative analysis of the theoretical spectra from different stellar atmosphere models. A visual inspection of the figure reveals that the most significant differences among the spectra, which affect the photoionization within the nebula, emerge beyond the ionization threshold of He~I ($\lambda$~504~\AA). The [WR] model exhibits a marked decline in flux, while the bb and p-p models reach their peaks closer to the He~II ionization threshold ($\lambda$~228~\AA). Beyond this point, an additional feature becomes evident: the bb model (dotted line) produces a much higher flux of UV photons compared to the p-p model (dot–dashed line), whereas the {[}WR{]} model contributes negligibly in this energy range. Examining the EUV region, we find that, similarly to the PN NGC~5315, the presence of a stellar wind plays a key role in attenuating the emerging radiation field, leading to a decrease in flux intensity within this high-energy domain, as illustrated by the {[}WR{]} model (solid line).

For the bb model, the logarithmic number of ionizing photons for H~I, He~I, and He~II are, respectively, 47.59, 47.18, and 45.72. For the p-p model, the corresponding values are 47.57, 47.28, and 44.47, while for the {[}WR{]} model, they are 47.58, 47.15, and 38.23. We conclude that the models converge to similar values for the number of photons energetic enough to ionize H~I but start to diverge significantly in the EUV region. The number of photons capable of ionizing He~II is drastically lower for the {[}WR{]} model, reinforcing our interpretation that the stellar wind increases opacity for EUV photons. Therefore, as in the case of NGC~5315, we conclude that including the stellar wind in the stellar atmosphere model has a significant impact on the shape of the emerging ionizing radiation field.

The best-fitting nebular parameters were found to be: a stellar luminosity of 5,000~L$_{\odot}$, a distance of 1,400~pc, an inner radius of $\log r_{0} = 17.80$~(cm), and a total hydrogen density of $\log n_{\mathrm{H}} = 4.0$. Model predictions were compared to the observed emission-line ratios reported by \citet{pottasch2003abundances}. The total extinction- and reddening-corrected H$_{\beta}$ flux (I$_{H_{\beta}}$) obtained by them was $1.73\times10^{-10}$~erg~cm$^{-2}$~s$^{-1}$, while our models predict values of $5.99\times10^{-11}$ (bb model), $3.92\times10^{-11}$ (p-p model), and $2.42\times10^{-10}$ ([WR] model) erg~cm$^{-2}$~s$^{-1}$. Among them, the {[}WR{]} model shows the best agreement with the observations.

\begin{figure*}
	\includegraphics[width=\textwidth]{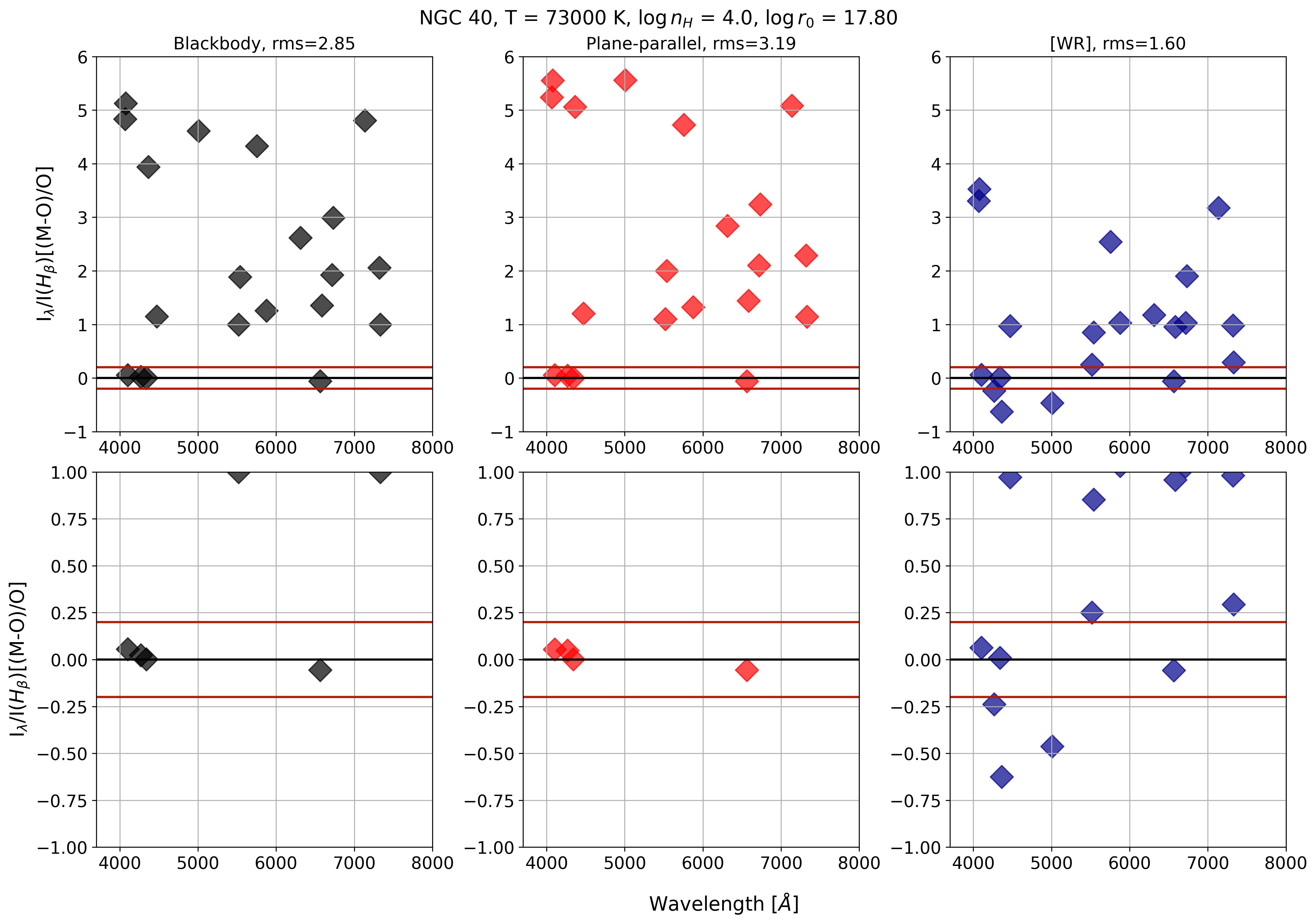}
    \caption{Same as Fig. \ref{fig:NGC5315lineratios}, but for NGC 40.}
    \label{fig:NGC40lineratios}
\end{figure*}

In Figure~\ref{fig:NGC40lineratios}, we present a comparison between the predicted and observed emission-line ratios, along with the root mean square (rms) deviation for each model (for completeness, see the numerical data in Appendix~\ref{ap:appendixA}). Despite several modeling attempts, the observed line ratios proved difficult to reproduce, with discrepancies reaching factors of 5 to 6 in some cases. This suggests that the assumptions in our \texttt{Cloudy} models may be somewhat simplified. Nevertheless, the {[}WR{]} model still outperforms both the bb and p-p models, with most of its predicted values deviating by less than 100\% (see lower panel of Fig.~\ref{fig:ngc6905lineratios}).

Our results indicate that reproducing the observed line ratios for this nebula requires more sophisticated and physically realistic modeling. None of the tested models could adequately reproduce the sulfur lines ([S~II] $\lambda\lambda$~4068, 4076, 6716, 6730), the nitrogen line [N~II] $\lambda$~6583, or the argon line [Ar~III] $\lambda$~7135, although the {[}WR{]} model achieved a slightly better match. Other studies have also encountered difficulties reproducing sulfur lines for this object (see, e.g., \citealt{henry2015co}), despite the [S~II]/H$_{\alpha}$ index showing no evidence of shock excitation (\citealt{daltabuit1976}).

Therefore, we conclude that for this PN, adopting a [WR] model for the central star still leads to an improvement in the photoionization results. However, it is clear that the models require further refinement and additional physical details to achieve a closer agreement with the observations. In the following section, we examine the case of a PN whose central star is even cooler.

\subsection{BD+30°3639}

The planetary nebula BD+30°3639, also known as Campbell's hydrogen star, is a low-excitation object (\citealt{aller1995nebular}) hosting a {[}WC9{]} type Wolf-Rayet central star (\citealt{acker2003quantitative}) with an estimated mass of 0.6~M$_{\odot}$ (\citealt{crowther2006}). Distance estimates for this nebula vary between 690 and 2800~pc (e.g., \citealt{pwa1986abundances,kawamura1996distances}). For consistency, in the present work we adopted a distance of 1200~pc, as used by \cite{marcolino2007detailed}, since their {[}WR{]} model was employed as the ionizing source in our photoionization simulations. Morphologically, BD+30°3639 exhibits a structure composed of ellipsoidal shells, with the central star approximately located at the geometric center of the nebula (see, for instance, \citealt{kastner2001discovery}).

\subsubsection{Results}

The central stellar model for BD+30°3639 was computed by \cite{marcolino2007detailed} using the CMFGEN code. The adopted model includes the following main parameters: an effective temperature of 47\,000~K (T${*}$), luminosity of 5000~L${\odot}$, stellar radius of 1.0~R${\odot}$, and surface gravity $\log g = 4.15$. The stellar wind is characterized by a mass-loss rate of $\log \dot{M} = -6.30$~(M${\odot}$\,yr$^{-1}$) and a terminal velocity of 700~km\,s$^{-1}$. We adopted a stellar mass of 0.60~M${\odot}$ from \citet{crowther2006ultraviolet}. For this nebula, the plane-parallel grid of models from \cite{rauch2003grid} does not provide an equivalent for the central star, since both its temperature and surface gravity are below the grid’s lower limits. Consequently, we compare only the blackbody (bb) and {[}WR{]} models in this case.

\begin{figure*}
	\includegraphics[width=\textwidth]{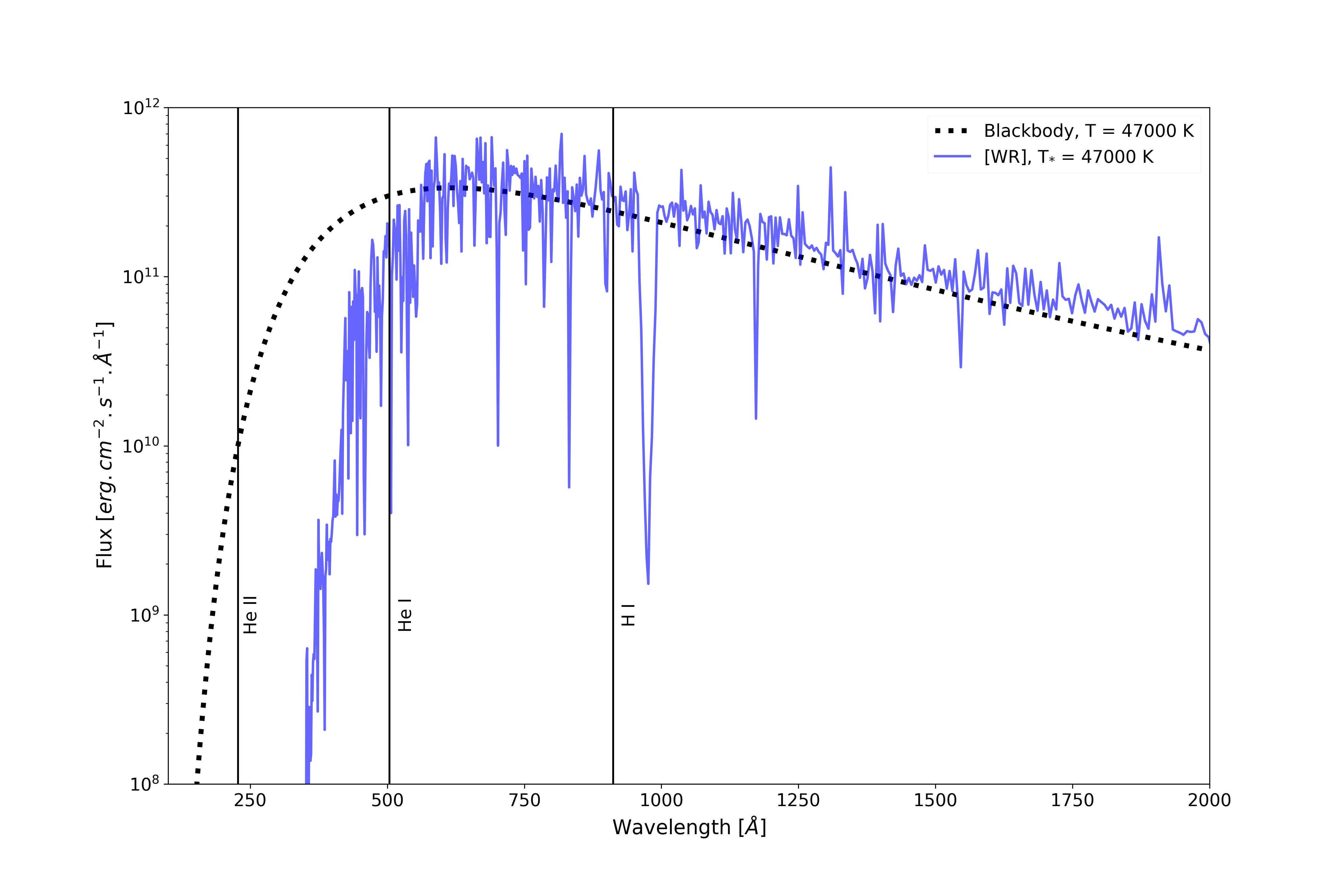}
    \caption{Same as Fig. \ref{fig:NGC5315fluxio}, but for BD+30 3639.}
    \label{fig:BD303639fluxio}
\end{figure*}

In Figure~\ref{fig:BD303639fluxio}, we present a comparison between the theoretical spectra of the stellar atmosphere models for the central star of BD+30°3639. A visual inspection of the figure reveals that the curves begin to diverge below the He~I ionization threshold, slightly earlier than in the previous cases involving late-type PNe. For this object, it is clear that the ionizing flux of the {[}WR{]} model undergoes a more pronounced decrease toward the EUV region, reinforcing our interpretation that the stellar wind increases the opacity of the emerging UV radiation field, particularly in the case of late-type central stars.

In the bb model, the logarithmic counts of ionizing photons for H~I, He~I, and He~II are 47.51, 46.73, and 44.14, respectively, while in the {[}WR{]} model they are 47.45, 45.97, and 32.24. From this comparison, we conclude that within this temperature range, the stellar wind exerts a stronger influence on the emerging flux than in the previous cases. Even the number of He~I ionizing photons shows a notable reduction, and the number of He~II photons drops drastically, while the number of H~I ionizing photons remains essentially ``wind-independent''. Therefore, we infer that the presence of the stellar wind introduces significant energetic differences in the emergent radiation field. To test whether these differences impact the predicted nebular line ratios, we employed these models as ionizing sources in our CLOUDY photoionization calculations and compared their predictions with observations.

The parameters of the photoionization models that yielded the best agreement between observed and theoretical line ratios are: T$_{*}$ = 47\,000~K, L = 5000~L$_{\odot}$, distance $d$ = 1200~pc, nebular radius $\log r_{0} = 17.27$~(cm), and total hydrogen density $\log n_{\mathrm{H}} = 4.2$. To quantify the agreement between our theoretical predictions and the observed values, we computed the rms between modelled and observed line ratios using the data from \cite{aller1995nebular}. The extinction- and reddening-corrected $H_\beta$ flux (I$_{H_{\beta}}$) reported by \cite{bernard2003abundances} is 3.10$\times$10$^{-10}$~erg\,cm$^{-2}$\,s$^{-1}$, while our predicted values are 1.79$\times$10$^{-11}$~erg\,cm$^{-2}$\,s$^{-1}$ for the bb model and 3.17$\times$10$^{-10}$~erg\,cm$^{-2}$\,s$^{-1}$ for the {[}WR{]} model. The {[}WR{]} prediction matches the observed flux both in order of magnitude and absolute value.

\begin{figure*}
	\includegraphics[width=\textwidth]{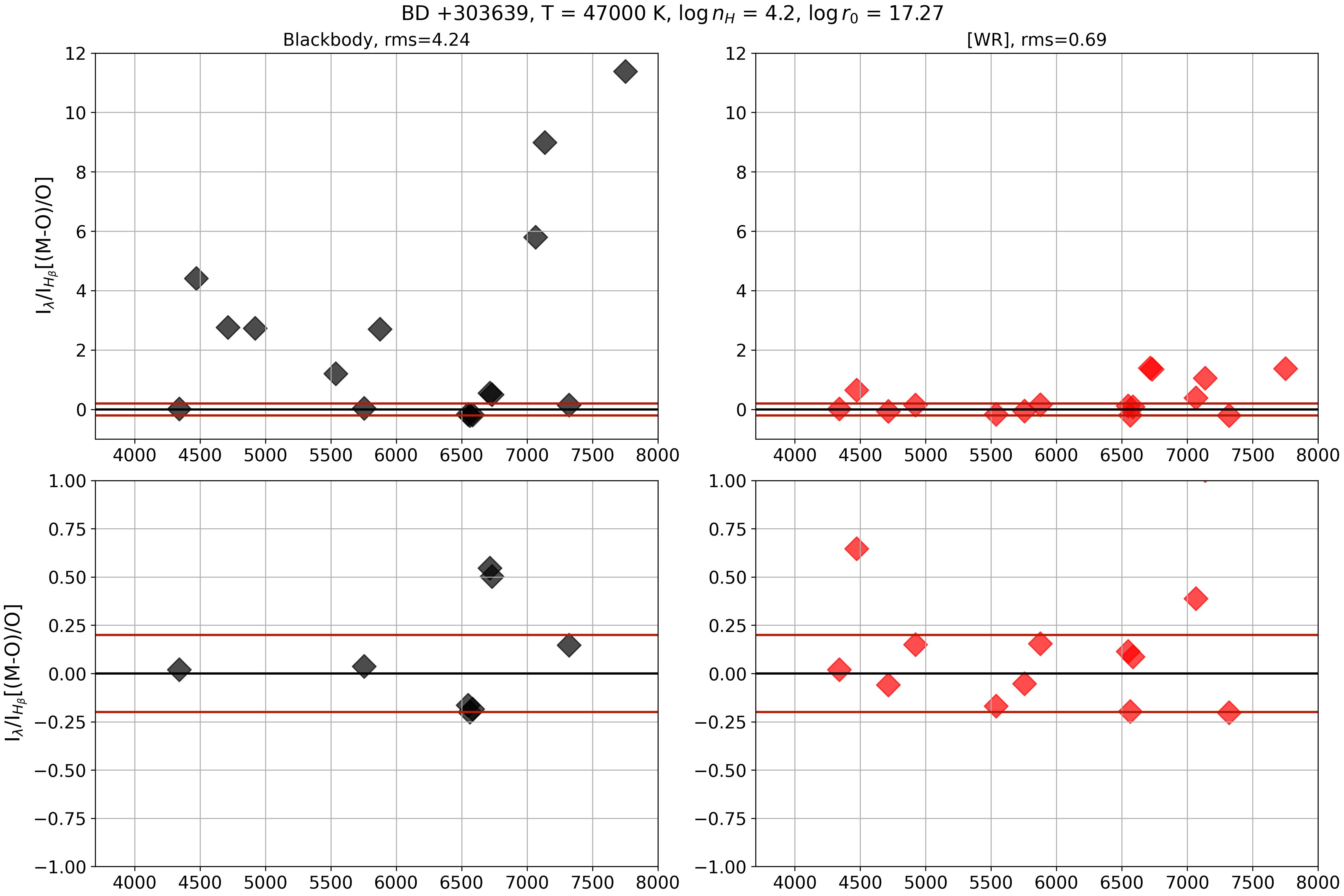}
    \caption{Same as Fig. \ref{fig:NGC5315lineratios}, but for BD+30 3639.}
    \label{fig:BD303639lineratios}
\end{figure*}

Figure~\ref{fig:BD303639lineratios} shows a comparison between the predicted and observed line ratios along with the rms values for each model. A visual inspection indicates that the {[}WR{]} model provides a much closer match between theory and observation. The rms values and individual line ratios confirm that, as in previous cases, the choice of stellar atmosphere model plays a decisive role in reproducing observed nebular spectra. Consequently, we conclude that, particularly for PNe with late-type central stars, accounting for the stellar wind improves the performance of photoionization models. This reinforces the need to consider wind effects in future studies.

In Table~\ref{tab:bd303639lineratios} of Appendix~\ref{ap:appendixA}, we list the modelled and observed line ratios for BD+30°3639. By inspecting the table, we note that neither model reproduces well the forbidden lines of argon, [Ar~III]~$\lambda\lambda$~7135.80, 7751.12~\AA, although the {[}WR{]} model provides a slightly better agreement. For the sulfur forbidden lines, [S~II]~$\lambda\lambda$~6716.47, 6730.85~\AA, the blackbody model performs somewhat better, though discrepancies exceed 60\%. For the remaining nebular lines, the {[}WR{]} model reproduces the observations within $\pm$20\%, which we consider satisfactory given the simplicity of our photoionization models.

\subsection{NGC 6905}

The PN NGC~6905 (also known as the Blue Flash Nebula) belongs to a small group of high-excitation planetary nebulae exhibiting prominent oxygen emission lines near $\lambda\lambda$3811 and 3834~\AA. This nebula is ionized by a [WC3] central star \citep{tylenda1993wolf}, with an estimated mass of 0.59~M$_{\odot}$ \citep{uzundag2021pulsating}. The star shows a mass-loss rate of $\log\dot{M} = -7.15$~(M$_{\odot}$~yr$^{-1}$), a terminal velocity $v_{\infty} = 1890$~km~s$^{-1}$, and an effective temperature T$_{*}$ between 140~kK and 165~kK \citep{pena1998galactic, marcolino2007detailed, keller2014uv}. For consistency, we adopted the temperature determined by \cite{marcolino2007detailed}, as their stellar atmosphere model was used in our analysis.

Distance estimates for this nebula range from 1730 to 1800~pc (see \citealt{keller2014uv} for a detailed compilation), but here we adopt the value of 1750~pc used by \cite{marcolino2007detailed}. The morphology of NGC~6905 is approximately elliptical, featuring a bright spheroidal shell with dimensions of $47^{\prime\prime} \times 34^{\prime\prime}$. Furthermore, the ellipsoidal axis appears to be inclined by about 60$^{\circ}$ relative to the line of sight \citep{cuesta1993spectroscopy}.

\subsubsection{Results}

We employed the [WR] model computed by \cite{marcolino2007detailed} using the CMFGEN code, along with equivalent stellar atmosphere models under plane-parallel and blackbody approximations, to analyze the ionizing flux and nebular properties. The model parameters are as follows: effective temperature T$_{*}$ = 146,000~K, luminosity $L = 5000$~L$_{\odot}$, stellar radius $R = 0.10$~R$_{\odot}$, and surface gravity $\log g = 6.2$. The stellar wind is characterized by a mass-loss rate $\log \dot{M} = -7.15$~(M$_{\odot}$~yr$^{-1}$) and a terminal velocity $v_{\infty} = 1890$~km~s$^{-1}$. We adopted a stellar mass of 0.59~M$_{\odot}$ \citep{uzundag2021pulsating}, consistent with the values discussed previously.

\begin{figure*}
	\includegraphics[width=\textwidth]{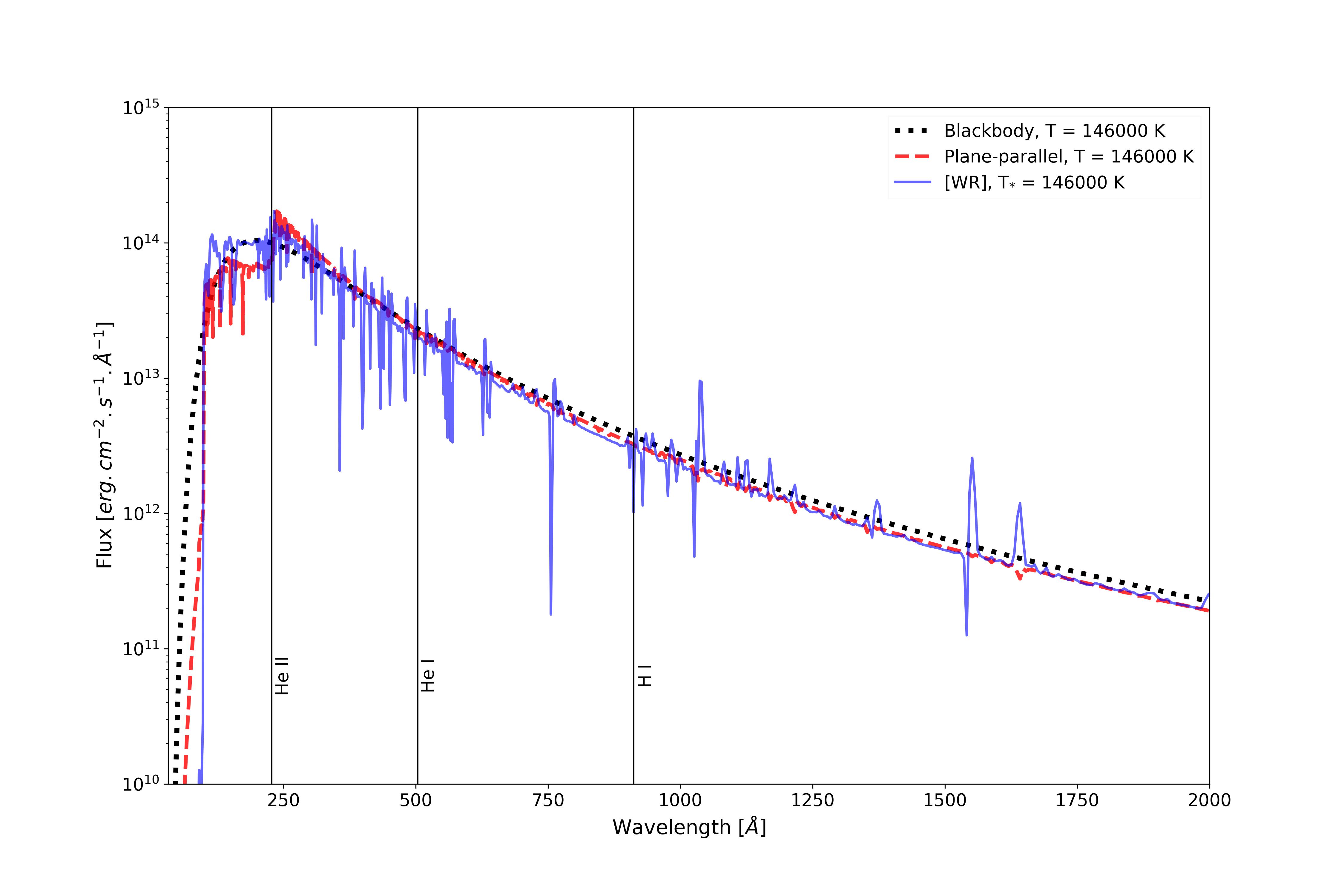}
    \caption{Same as Fig. \ref{fig:NGC5315fluxio}, but for NGC 6905.}
    \label{fig:ngc6905fluxio}
\end{figure*}

Figure~\ref{fig:ngc6905fluxio} compares the theoretical spectra of the stellar atmosphere models. Differences among the models become apparent above the He~II ionization threshold ($\lambda$~228~\AA), similar to what was observed for the hottest models (T$_{*} \geq 150$~kK) discussed in Section~\ref{sec:sec2}. In this region, the plane-parallel model shows a slightly steeper flux decline, implying a lower production of He~II-ionizing photons. This suggests that, for NGC~6905, the stellar wind has a weaker influence on the radiation field compared to cooler [WR] stars. Consequently, the significance of stellar winds as an additional opacity source appears to be temperature dependent, important for late-type [WR] stars, but less relevant or negligible for early-type central stars.

For the bb model, the logarithmic number of ionizing photons for H~I, He~I, and He~II are 47.47, 47.33, and 46.76, respectively. For the plane-parallel (p-p) model, these values are 47.48, 47.35, and 46.62, while for the [WR] model they are 47.45, 47.33, and 46.75. From these results, we note no significant differences among the models, except that the plane-parallel model produces slightly fewer He~II-ionizing photons, consistent with the spectral behavior shown in Figure~\ref{fig:ngc6905fluxio}. Based on the radiation field analysis and ionizing photon counts, we conclude that the choice of stellar atmosphere model does not have a strong influence on the photoionization line-ratio predictions, in contrast to the cooler cases discussed earlier. This suggests that, within this temperature range, the stellar wind has little to no effect on the opacity for UV photons. Thus, early-type [WR] PNe may be modeled without significant concern regarding the stellar atmosphere adopted. This conclusion is consistent with the analysis presented in Section~\ref{sec:sec2} for similar temperature regimes.

The best-fitting photoionization model parameters reproducing the observed nebular properties are a stellar luminosity of 5000~L$_{\odot}$, a distance of 1750~pc, an inner nebular radius of $\log r_{0} = 17.45$~(cm), and a total hydrogen density of $\log n_{\mathrm{H}} = 3.4$. The observed emission-line ratios used for comparison were taken from \citet{gomez2022planetary}, who obtained spectra for eight distinct regions across the nebula (see their Figure~1) and modeled them using CLOUDY. In our analysis, we compared our results to the region labeled A1 in their work. The emission lines used in our comparison are among the most intense in the optical spectrum and are commonly adopted to optimize photoionization models (e.g., \citealt{bohigas2008photoionization}). We also tested the effect of varying the slit parameters in CLOUDY, as performed by \citet{gomez2022planetary}, but found no significant improvement in the fits. The total extinction- and reddening-corrected $H_{\beta}$ flux obtained by them is 1.29$\times$10$^{-11}$~erg~cm$^{-2}$~s$^{-1}$, while our theoretical predictions are 7.38$\times$10$^{-10}$~erg~cm$^{-2}$~s$^{-1}$ (bb model), 5.45$\times$10$^{-10}$~erg~cm$^{-2}$~s$^{-1}$ (p-p model), and 9.34$\times$10$^{-10}$~erg~cm$^{-2}$~s$^{-1}$ ([WR] model). In all cases, our predicted values are higher than the observed one; increasing the adopted distance within the literature range did not significantly change this discrepancy.

\begin{figure*}
	\includegraphics[width=\textwidth]{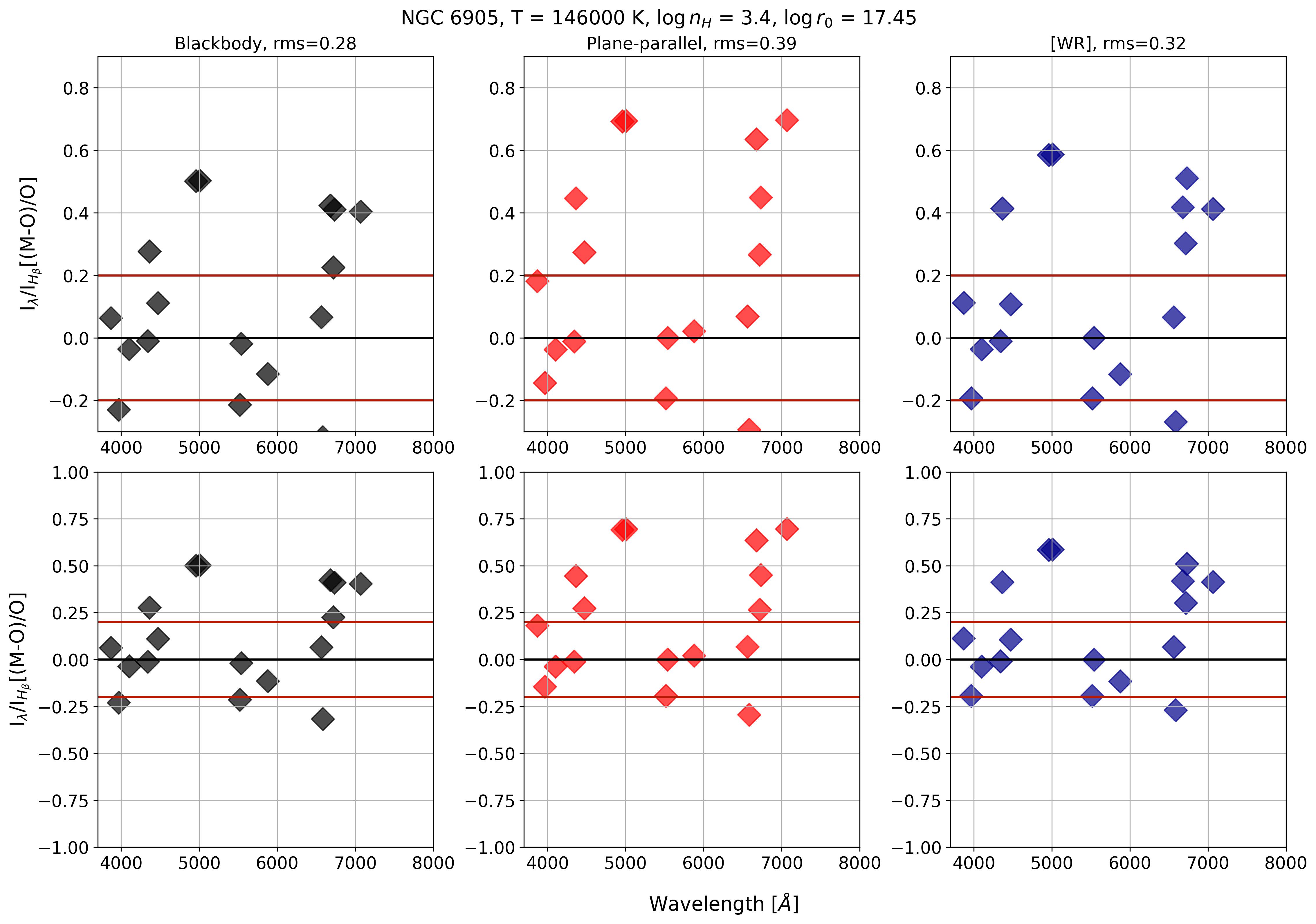}
    \caption{Same as Fig. \ref{fig:NGC5315lineratios}, but for NGC 6905.}
    \label{fig:ngc6905lineratios}
\end{figure*}

Figure~\ref{fig:ngc6905lineratios} presents a comparison between the observed and predicted line ratios, along with the rms values for each model. The bb model produced the lowest rms, although the [WR] model yielded a similar result. The p-p model, however, overestimates the intensity of several lines, such as [O~III]~$\lambda$5007~\AA\ and [Ne~III]~$\lambda$3868~\AA. Overall, none of the models reproduce the observed spectrum within the $\pm$20\% uncertainty level, though the rms values indicate that adjusting nebular parameters could improve the fits regardless of the adopted stellar atmosphere. For this PN, our results suggest that in early-type [WR] objects, the stellar wind does not play a significant role as an additional opacity source, and the choice of the stellar atmosphere model may be a minor concern. All models failed to reproduce the forbidden lines of oxygen and nitrogen (e.g., [O~III]~$\lambda$5007~\AA, [N~II]~$\lambda$6583~\AA) within $\pm$20\% (numerical values are provided in Appendix~\ref{ap:appendixA}), while other transitions were well matched. These findings are consistent with \citet{gomez2022planetary}, and a more detailed modeling effort would be required to improve the prediction of these key cooling lines in planetary nebulae.

\subsection{NGC 2867}

The PN NGC~2867 is a high-excitation nebula \citep{pena2017kinematic}, ionized by a [WO2] central star \citep{acker2003quantitative} with an estimated mass of 0.57~M$_{\odot}$. The central star is undergoing mass loss at a rate of 3.0$\times$10$^{-8}$~M$_{\odot}$~yr$^{-1}$, with a terminal wind velocity of 2000~km~s$^{-1}$ and an estimated temperature (T$_{*}$) of 165,000~K \citep{keller2014uv}. Reported distance estimates for this object range from 1800 to 2200~pc \citep[e.g.,][]{keller2014uv,pena2017kinematic}. In this work, we adopted a distance of 1840~pc, corresponding to the value derived by \citet{keller2014uv} from their stellar atmosphere modeling, whose model we also employed for our flux and nebular analyses.

Morphologically, NGC~2867 presents an elliptical structure with an estimated angular size of $14.4^{\prime\prime}\times13.9^{\prime\prime}$ \citep{tylenda2003angular}. The nebula exhibits a complex shell morphology, featuring multiple layers and an overall diameter of approximately $25^{\prime\prime}$ \citep{pena1998galactic}.

\subsubsection{Results}
\label{sec:results}

For this object, the original stellar model computed by \citet{keller2014uv} was not available. Nevertheless, using their reported parameters as reference, we selected a [WR] model from the grid of expanding atmosphere models by \citet{keller2011new}, whose parameters most closely matched those derived for this nebula. The adopted model includes the following main parameters: an effective temperature of 165,000~K (T${*}$), luminosity of 5800~L${\odot}$, stellar radius of 0.09~R${\odot}$, and surface gravity of $\log g = 6.3$. The stellar wind is described by a mass-loss rate of $\log \dot{M} = -7.0$~(M${\odot}$~yr$^{-1}$) and a terminal velocity of 2000~km~s$^{-1}$. For this source, we adopted a stellar mass of 0.59~M${\odot}$ \citep{keller2011new}.

\begin{figure*}
	\includegraphics[width=\textwidth]{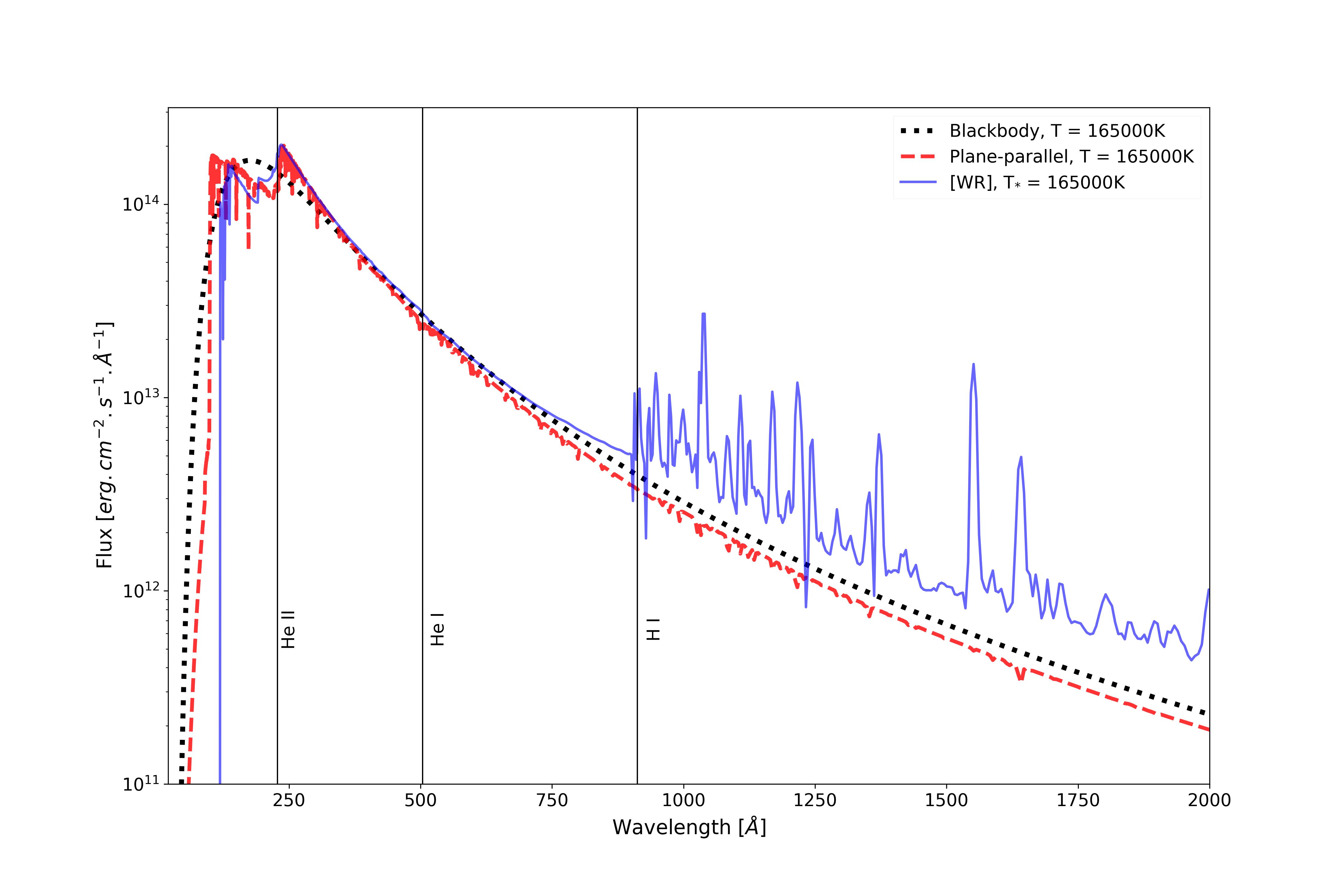}
    \caption{Same as Fig. \ref{fig:NGC5315fluxio}, but for NGC 2867.}
    \label{fig:ngc2867fluxio}
\end{figure*}

In Figure~\ref{fig:ngc2867fluxio}, we present a comparison between the theoretical spectra of the stellar atmosphere models. From this figure, it can be noted that, as in the previous case, the spectral energy distributions begin to diverge significantly beyond the He~II ionization threshold ($\lambda = 228$~\AA). However, near this threshold, both the [WR] and p-p models show a slight enhancement compared to the bb model. At shorter wavelengths, the differences persist ,  the [WR] model declines more rapidly, while the p-p model exhibits the highest flux peak. This qualitative comparison suggests, similarly to the case of early-type CSPNe analyzed earlier, that the contribution of the stellar wind as an additional opacity source for UV photons appears to be small or negligible, and its impact is strongly temperature dependent.

For the bb model, the number of ionizing photons for H~I, He~I, and He~II are, respectively, 47.49, 47.38, and 46.90. For the p-p model, the corresponding values are 47.49, 47.39, and 46.85, while for the [WR] model they are 47.50, 47.39, and 46.77. The number of ionizing photons for H~I remains essentially the same across all models. For He~I, the [WR] model produces slightly more ionizing photons, followed by the bb and p-p models, consistent with their spectral energy distributions (see Fig.~\ref{fig:ngc2867fluxio}). For He~II, the p-p model yields the highest number of ionizing photons, followed by the bb and [WR] models. This result highlights how even subtle variations in the radiation field can lead to measurable differences in the predicted ionizing photon fluxes. Moreover, it suggests that for this central star, the stellar wind still contributes to the opacity affecting the EUV radiation, albeit on a smaller scale compared to late-type stars.

The best-fitting nebular parameters derived for this object are a stellar luminosity of 5000~L${\odot}$, a distance of 1840~pc, an inner nebular radius of $\log r_{0} = 17.00$~(cm), and a total hydrogen density of $\log n_{\mathrm{H}} = 4.0$. The model predictions were compared against the observed emission-line ratios reported by \citet{aller1981analysis}. The total extinction- and reddening-corrected H$\beta$ flux (I$_{H_{\beta}}$) provided by \citet{acker1992strasbourg} is 5.89$\times$10$^{-11}$~erg~cm$^{-2}$~s$^{-1}$, while our theoretical predictions are 7.38$\times$10$^{-10}$ (bb model), 5.45$\times$10$^{-10}$ (p-p model), and 9.34$\times$10$^{-10}$ ([WR] model)~erg~cm$^{-2}$~s$^{-1}$, approximately one order of magnitude higher than the observed value.

\begin{figure*}
	\includegraphics[width=\textwidth]{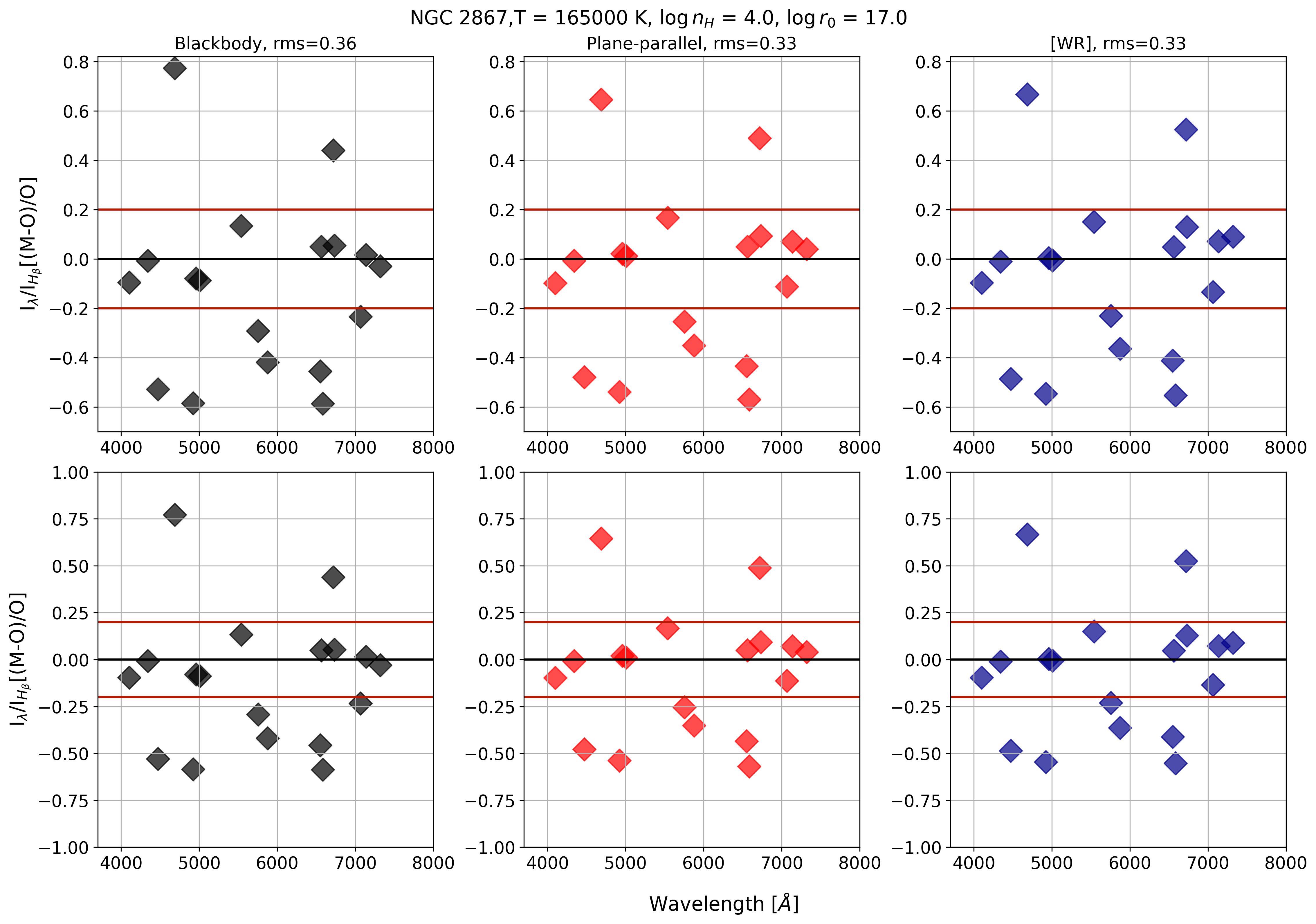}
    \caption{Same as Fig. \ref{fig:NGC5315lineratios}, but for NGC 2867.}
    \label{fig:ngc2867lineratios}
\end{figure*}

Figure~\ref{fig:ngc2867lineratios} presents a comparison between the predicted and observed line ratios along with the rms for each model. A visual inspection reveals that the helium lines contribute most to the rms increase, while the majority of forbidden metal lines remain within $\pm$20\% of the observed values, except for the nitrogen lines discussed previously (complete numerical data are available in Appendix~\ref{ap:appendixA}). Overall, the three ionizing source models yield similar rms values, indicating comparable performance in reproducing the observed line ratios. Small refinements to the input parameters in CLOUDY could further improve the agreement among all models.

Our models successfully reproduced most of the observed emission lines using the adopted approach. However, none of the models provided satisfactory fits for the [N~II] $\lambda\lambda$6548.30 and 6583.41 lines. This limitation likely arises from the simplified nature of our photoionization models and possibly from the assumed standard PN abundances not accurately reflecting the actual nebular composition. In addition, none of the models accurately reproduced the observed He~I and He~II line ratios (e.g., He~II~$\lambda$4472, He~II~$\lambda$4686, He~I~$\lambda$4922), which were consistently predicted to be stronger than observed. Given that the ratio He~II~$\lambda$4686/H$\beta$ is highly sensitive to the central star temperature \citep{kaler1989central}, this discrepancy could imply that a cooler central star temperature might be required to better match the observations. For instance, \citet{deguchi1997iau} derived a temperature of 141,000~K for this central star. However, adopting a different stellar temperature would necessitate recalibrating the nebular parameters to preserve the fits of the other lines, which falls outside the scope of the present study.

Our results for this planetary nebula indicate that the choice of stellar atmosphere model has only a minor impact on the resulting photoionization model. Similarly, the stellar wind does not appear to contribute significantly to the opacity of UV photons, consistent with our findings for NGC 6905. These results suggest that the influence of stellar winds on photoionization models may depend on temperature, playing a more prominent role in late-type [WR] central stars and becoming less relevant in early-type ones. However, additional studies are required to verify this potential trend.

\section{Limitations of Early Atmosphere Models in Photoionization Studies}
\label{sec:limitations}

The methodology described in Section~\ref{sec:sec3} was similarly applied by \citealt{howard1997detailed}. These authors computed nebular line ratios predicted by photoionization models using different central ionizing sources. In addition to the blackbody approximation, they employed plane-parallel LTE models from \citealt{kurucz1991precision}, non-LTE CNO line-blanketed models from \cite{werner1991}, non-LTE hydrogen models from \cite{clegg1987non}, and non-LTE plane-parallel models from \cite{husfeld1984non}. Their sample included nine planetary nebulae (PNe) from the Galactic halo, at least two of which show evidence of a moderate stellar wind (DdDm-1 and K~648), and one (M~2-49) hosting a confirmed hydrogen-deficient central star (see, e.g., \citealt{pena1992ultraviolet}).

Figure~A1 of their paper presents all line ratios predicted by the different models compared to the observed values for each nebula. Overall, they concluded that the differences in predicted nebular lines using various ionizing sources were not significant. In their words, ``nebular line strengths are relatively insensitive to atmospheric details; thus blackbody spectra are suitable for central star continua.'' In other words, the choice of ionizing source appeared to have little impact on the results.

However, compared to the CMFGEN-based atmosphere models, those used by \cite{howard1997detailed} are relatively outdated and present key limitations. Several models rely on LTE, which is inadequate for CSPNe, and they neglect stellar wind effects in many cases. Additionally, only moderate line blanketing is included, with limited metallic content or none at all. These shortcomings alone cast doubt on the reliability of their predicted line ratios. A closer inspection of their Figure~A1 reveals that, contrary to their conclusion, line strengths can, in fact, be sensitive to the choice of ionizing continuum. For example, for PN~006$-$41.9, depending on the model used, the discrepancy between the predicted and observed [N~II]~$\lambda$6584 or [O~II]~$\lambda$3727 line ratios varies from nearly 0 to about $\pm$0.3~dex (a factor of $\sim$2).

In summary, our calculations, which employ modern stellar atmosphere models, are more consistent and reliable than those presented by \cite{howard1997detailed}. Our results demonstrate quantitatively that nebular line ratios can be highly sensitive to the adopted ionizing source. In some cases, the use of blackbody continua can yield predictions that deviate substantially from observations (e.g., BD+30°3639; see Figure~\ref{fig:BD303639lineratios}).

\section{Discussion and conclusions}
\label{sec:conclusions}

We analyzed a set of central star atmosphere models based on their theoretical spectra and the corresponding number of ionizing photons for H~I, He~I, and He~II. Additionally, we examined the impact of different radiation field models on the predicted nebular line ratios using photoionization simulations for a sample of five planetary nebulae hosting [WR]-type central stars: NGC~5315, NGC~6905, NGC~2867, NGC~40, and BD+30°3639.

Our analysis demonstrates that, in certain cases, the stellar wind significantly influences the emerging radiation field of the central star; however, this effect appears to be temperature-dependent. For models with lower effective temperatures (T$_{*} \lesssim 100$~kK), the wind drastically alters the extreme ultraviolet (EUV) portion of the spectrum (see, for instance, Figure~\ref{fig:fluxA50} in Section~\ref{sec:sec2}), leading to a reduced number of helium-ionizing photons. This suggests that the stellar wind acts as an additional source of opacity, attenuating UV photons. In contrast, for hotter stars (T$_{*} \gtrsim 100$~kK), the impact of the wind becomes less pronounced. In such models, the gas is highly ionized, becoming more transparent to the EUV flux when compared with cooler stars (as the ground levels of important transitions become much less populated).

In Section~\ref{sec:sec3}, for each PN in our sample, we applied the same methodology as in Section~\ref{sec:sec2} and then used these atmosphere models to build our photoionization models. The results from comparing the theoretical spectra and the number of ionizing photons for each model confirm our earlier findings: in early-type [WR] stars (T$_{*} \gtrsim 100$~kK), such as NGC~6905, the radiation field is less sensitive to the choice of atmosphere model. However, in late-type [WR] stars (T$_{*} \lesssim 100$~kK), such as NGC~5315 and BD+30°3639, the presence of a stellar wind has a more substantial impact, reinforcing the temperature dependency of the wind's role as a source of opacity for the emerging photons.

Our nebular analysis indicates that stellar winds play a critical role in shaping the ionizing spectrum in PNe with late-type [WR] central stars (e.g., NGC~40, NGC~5315, BD+30°3639), which typically have T$_{*} < 80\,000$~K. For early-type [WR] stars (e.g., NGC~2867, NGC~6905), the effect of the wind is comparatively minor. Nevertheless, to further validate these trends, a more extensive investigation with a larger and more diverse sample of PNe, alongside more sophisticated stellar atmosphere and nebular models, is necessary.

Constructing photoionization models requires considerable computational effort, including some optimization tests, and selecting an appropriate set of input parameters for each nebula was often a demanding task. For certain cases, such as NGC~40, even after exhaustive optimization efforts, the standard 1D photoionization models failed to fully reproduce the observed spectra, suggesting that more complex physics is at play.

We acknowledge that factors such as nebular expansion, density fluctuations, geometry, and other physical conditions (e.g., shocks) influence the predicted line ratios. Nevertheless, our simplified photoionization models demonstrate that [WR] stellar atmospheres should be employed in more sophisticated modeling of PNe hosting such central stars. To further improve the performance of our models, it would be necessary to account for additional ingredients not included in this study, such as more realistic nebular geometries, the presence of dust, and refined chemical abundance distributions. A promising avenue would be the development of multi-dimensional models for each object, coupled with a broader grid of stellar input spectra.

\section*{Acknowledgments}

L.V.C, C.M, and W.M. thanks Coordenação de Aperfeiçoamento de Pessoal de Nível Superior (CAPES) for financial support and the Valongo Observatory for providing the facilities and tools necessary for the achievements of this work.

\bibliographystyle{apsrev4-1}

\bibliography{References}

\begin{appendix}

\section{Appendix A: Line ratios}
\label{ap:appendixA}

\begin{table*}[!h]
\centering
\caption[NGC 5315: Nebular models results]{NGC 5315: Nebular models results}
\label{tab:razaolinhasNGC5315}
\begin{tabular}{c|ccccccc}
\hline
\textbf{Species} & \textbf{$\lambda_{0}$ (\AA)} & \textbf{Observed (I$_{\lambda}$/I$_{H_{\beta}}$)} & \textbf{Blackbody (I$_{\lambda}$/I$_{H_{\beta}}$)} & \textbf{Plane-parallel (I$_{\lambda}$/I$_{H_{\beta}}$)} & \textbf{{[}WR{]} (I$_{\lambda}$/I$_{H_{\beta}}$)} \\
\hline
{[}S II{]} & 4068.7 & 5.590 & 8.87 & 9.49 & 7.65 \\
H I & 4101 & 26.110 & 26.18 & 25.20 & 25.39 \\ 
H I & 4340 & 47.840 & 46.14 & 46.14 & 46.43 \\
{[}O III{]} & 4363.2 & 4.39 & 8.33 & 9.31 & 6.200 \\
He I & 4471.5 & 6.42 & 5.18 & 5.24 & 5.16 \\
{[}Ar IV{]} & 4711.3 & 0.060 & 0.88 & 0.83 & 0.11 \\
He II & 4713.1 & 0.840 & 0.82 & 0.85 & 0.82 \\
{[}Ar IV{]} & 4740.1 & 0.26 & 1.08 & 1.02 & 0.14 \\
{[}O III{]} & 4958.9 & 282.00 & 348.48 & 374.71 & 281.42 \\
{[}O III{]} & 5006.9 & 842.90 & 1044.89 & 1120.34 & 841.29 \\
{[}N II{]} & 5754.60 & 4.46 & 5.89 & 6.14 & 5.56 \\
He I & 5875.6 & 19.35 & 15.11 & 16.26 & 15.77 \\
{[}O I{]} & 6300.3 & 4.10 & 8.97 & 8.94 & 6.24 \\ 
{[}S III{]} & 6312.1 & 3.36 & 3.99 & 4.14 & 4 \\ 
{[}O I{]} & 6363.8 & 1.42 & 2.87 & 3.05 & 2 \\ 
{[}N II{]} & 6548.1 & 49.30 & 62.83 & 63.83 & 61.09 \\ 
H I & 6562.82 & 294.8 & 295.20 & 295.37 & 293.65 \\ 
{[}N II{]} & 6583.5 & 151.70 & 185.42 & 188.38 & 180.26 \\
He I & 6678 & 5.16 & 4.07 & 4.2 & 4.10 \\ 
{[}S II{]} & 6716.4 & 3.11 & 3.87 & 4.1 & 3.13 \\
{[}S II{]} & 6730.8 & 6.55 & 8.31 & 8.81 & 6.78 \\
He I & 7065 & 9.25 & 10.05 & 10.58 & 10.01 \\
{[}O II{]} & 7319.5 & 4.34 & 13.28 & 13.63 & 15.81 \\
{[}O II{]}& 7330.2 & 3.15 & 7.15 & 7.34 & 8.51 \\
\hline
        
\end{tabular} \\ 
\flushleft
\footnotesize{From left to right, the columns represent: The species names (atom/ion), their laboratory wavelengths, the observed nebular line ratios from \cite{peimbert2004physical}, and the last three columns are the theoretical line ratios predicted by the photoionization models using the indicated stellar atmosphere approximations.}
\end{table*}

\begin{table*}
\centering
\caption[NGC 40: Nebular models results]{NGC 40: Nebular models results}
\label{tab:razoeslinhasNGC40}
\begin{tabular}{c|ccccccc}
\hline
\textbf{Species} & \textbf{$\lambda_{0}$ (\AA)}  & \textbf{Observed (I$_{\lambda}$/I$_{H_{\beta}}$)} & \textbf{blackbody (I$_{\lambda}$/I$_{H_{\beta}}$)} & \textbf{plane-parallel (I$_{\lambda}$/I$_{H_{\beta}}$)} & \textbf{{[}WR{]} (I$_{\lambda}$/I$_{H_{\beta}}$)} \\
\hline
{[}S II{]} & 4068.60 & 3.93 & 22.93 & 24.53 & 16.92 \\
{[}S II{]} & 4076.35 & 1.21 & 7.41 & 7.93 & 5.48 \\
H I & 4101.74 & 24.51 & 25.87 & 25.85 & 26.07 \\
C II & 4267.15 & 0.42 & 0.43 & 0.44 & 0.32 \\
H I & 4340.47 & 46.82 & 46.9 & 46.91 & 47.24 \\
{[}O III{]} & 4363.21 & 0.16 & 0.79 & 0.97 & 0.06 \\
He I & 4471.50 & 2.43 & 5.22 & 5.36 & 4.79 \\
{[}O III{]} & 5006.84 & 24.86 & 139.51 & 163.01 & 13.36 \\
{[}Cl III{]} & 5517.71 & 0.20 & 0.4 & 0.42 & 0.25 \\
{[}Cl III{]} & 5537.88 & 0.27 & 0.78 & 0.81 & 0.5 \\
{[}N II{]} & 5754.64 & 2.14 & 11.4 & 12.25 & 7.58 \\
He I & 5875.64 & 6.53 & 14.74 & 15.18 & 13.27 \\
{[}S III{]} & 6312.10 & 0.50 & 1.81 & 1.92 & 1.09 \\
H I & 6562.82 & 300.58 & 283.8 & 283.91 & 283.25 \\
{[}N II{]} & 6583.41 & 283.24 & 667.7 & 692.62 & 554.61 \\
{[}S II{]} & 6716.47 & 10.40 & 30.42 & 32.25 & 21.12 \\
{[}S II{]} & 6730.85 & 14.45 & 57.71 & 61.25 & 41.95 \\
{[}Ar III{]} & 7135.80 & 4.62 & 26.83 & 28.09 & 19.29 \\
{[}O II{]} & 7319.99 & 6.59 & 20.18 & 21.66 & 13.06 \\
{[}O II{]} & 7330.73 & 5.43 & 10.86 & 11.66 & 7.03 \\
\hline
\end{tabular}\\ 
\flushleft
\footnotesize{From left to right the columns represent: The species names (atom/ion), their laboratory wavelengths, the observed nebular line ratios from \cite{pottasch2003abundances}, and the three last columns are the theoretical line ratios predicted by the photoionization models using the indicated stellar atmosphere approximations.}
\end{table*}

\begin{table*}
\centering
\caption[BD+303639: Nebular models results]{BD +303639: Nebular models results}
\label{tab:bd303639lineratios}
\begin{tabular}{c|ccccccc}
\hline
\textbf{Species} & \textbf{$\lambda_{0}$ (\AA)} & \textbf{Observed (I$_{\lambda}$/I$_{H_{\beta}}$)} & \textbf{blackbody (I$_{\lambda}$/I$_{H_{\beta}}$)} & \textbf{{[}WR{]} (I$_{\lambda}$/I$_{H_{\beta}}$)} \\
\hline
 H I & 4340.47 & 45.56 & 46.49 & 46.46 \\
He I & 4471.50 & 0.90 & 4.83 & 1.47 \\
He I & 4713.17 & 0.17 & 0.64 & 0.16 \\
He I & 4921.93 & 0.35 & 1.30 & 0.40 \\
{[}Cl III{]} & 5537.88 & 0.35 & 0.77 & 0.29 \\
{[}N II{]} & 5754.64 & 4.63 & 4.80 & 4.39 \\
He I & 5875.67 & 3.56 & 13.19 & 4.11 \\
{[}N II{]} & 6548.03 & 119.87 & 100.30 & 133.61 \\
H I & 6562.82 & 362.15 & 289.72 & 291.04 \\
{[}N II{]} & 6583.41 & 362.87 & 295.88 & 394.12 \\
{[}S II{]} & 6716.47 & 4.04 & 6.25 & 9.69 \\
{[}S II{]} & 6730.85 & 8.71 & 13.09 & 20.50 \\
He I & 7065.25 & 1.02 & 6.91 & 1.41 \\
{[}Ar III{]} & 7135.80 & 2.19 & 21.90 & 4.50 \\
{[}O II{]} & 7319.99 & 9.17 & 10.51 & 7.31 \\
{[}Ar III{]} & 7751.12 & 0.41 & 5.05 & 0.97 \\
\hline
\end{tabular} \\ 
\flushleft
\footnotesize{From left to right the columns represent: The species names (atom/ion), their laboratory wavelengths, the observed nebular line ratios from \cite{aller1995nebular}, and the three last columns are the theoretical line ratios predicted by the photoionization models using the indicated stellar atmosphere approximations.}
\end{table*}

\begin{table*}
\centering
\caption[NGC 6905: Nebular models results]{NGC 6905: Nebular models results}
\label{tab:linerationsNGC6905}
\begin{tabular}{c|ccccccc}
\hline
\textbf{Species} & \textbf{$\lambda_{0}$ (\AA)} & \textbf{Observed (I$_{\lambda}$/I$_{H_{\beta}}$)} & \textbf{blackbody (I$_{\lambda}$/I$_{H_{\beta}}$)} & \textbf{plane-parallel (I$_{\lambda}$/I$_{H_{\beta}}$)} & \textbf{{[}WR{]} (I$_{\lambda}$/I$_{H_{\beta}}$)} \\
\hline
{[}Ne III{]} & 3868.75 & 108.00 & 114.82 & 127.61 & 120.17 \\
{[}NeIII{]} & 3967.46 & 45.60 & 35.13 & 39.05 & 36.78 \\
H I & 4101.74 & 25.70 & 24.78 & 24.74 & 24.76 \\
H I & 4340.47 & 46.20 & 45.71 & 45.69 & 45.74 \\
{[}O III{]} & 4363.21 & 13.80 & 17.62 & 19.96 & 19.51 \\
He I & 4471.50 & 2.70 & 3.00 & 3.44 & 2.99 \\
{[}O III{]} & 4958.91 & 347.00 & 520.96 & 587.17 & 549.86 \\
{[}O III{]} & 5006.84 & 1040.00 & 1559.09 & 1757.30 & 1645.54 \\
{[}Cl III{]} & 5517.71 & 1.50 & 1.18 & 1.21 & 1.21 \\
{[}Cl III{]} & 5537.88 & 1.10  & 1.08 & 1.10 & 1.10 \\
He I & 5875.67 & 9.90 & 8.76 & 10.11 & 8.75 \\
H I & 6562.82 & 283.00 & 301.85 & 302.33 & 301.72 \\
{[}N II{]} & 6583.41 & 187.00 & 127.68 & 132.11 & 136.76 \\
He I & 6678.15 & 1.70 & 2.42 & 2.78 & 2.41 \\
{[}S II{]} & 6716.47 & 17.40 & 21.33 & 22.04 & 22.66 \\
{[}S II{]} & 6730.85 & 15.50 & 21.86 & 22.47 & 23.42 \\
He I & 7065.25 & 2.40 & 3.37 & 4.07 & 3.39 \\
\hline
\end{tabular}\\ 
\flushleft
\footnotesize{From left to right, the columns represent: The species names (atom/ion), their laboratory wavelengths, the observed nebular line ratios from \cite{gomez2022planetary}, and the last three columns are the theoretical line ratios predicted by the photoionization models using the indicated stellar atmosphere approximations.}
\end{table*}

\begin{table*}
\centering
\caption[NGC 2867: Nebular models results]{NGC 2867: Nebular models results}
\label{tab:lineratioNGC2867}
\begin{tabular}{c|ccccccc}
\hline
\textbf{Species} & \textbf{$\lambda_{0}$ (\AA)} & \textbf{Observed (I$_{\lambda}$/I$_{H_{\beta}}$)} & \textbf{blackbody (I$_{\lambda}$/I$_{H_{\beta}}$)} & \textbf{plane-parallel (I$_{\lambda}$/I$_{H_{\beta}}$)} & \textbf{{[}WR{]} (I$_{\lambda}$/I$_{H_{\beta}}$)} \\
\hline
H I & 4101.74 & 26.60 & 24.07 & 24.01 & 24.03 \\
H I & 4340.47 & 45.20 & 44.87 & 44.87 & 44.71 \\
He I & 4471.50 & 4.20 & 1.98 & 2.19 & 2.16 \\
He II & 4685.64 & 31.50 & 55.82 & 51.84 & 52.49 \\
He I & 4921.93 & 1.30 & 0.54 & 0.60 & 0.59 \\
{[}O III{]} & 4958.91 & 470.00 & 432.79 & 479.77 & 471.46 \\
{[}O III{]} & 5006.84 & 1422 & 1297.16 & 1438.03 & 1413.08 \\
{[}Cl III{]} & 5537.88 & 0.60 & 0.68 & 0.70 & 0.69 \\
{[}N II{]} & 5754.64 & 1.30 & 0.92 & 0.97 & 1.00 \\
He I & 5875.67 & 10.80 & 6.28 & 7.01 & 6.87 \\
{[}N II{]} & 6548.03 & 26.00 & 14.15 & 14.71 & 15.29 \\
H I & 6562.82 & 301.00 & 315.73 & 315.91 & 315.6 \\
{[}N II{]} & 6583.41 & 101.00 & 41.79 & 43.46 & 45.18 \\
{[}S II{]} & 6716.47 & 5.30 & 7.63 & 7.89 & 8.08 \\
{[}S II{]} & 6730.85 & 7.30 & 7.69 & 7.98 & 8.24 \\
He I & 7065.25 & 4.10 & 3.14 & 3.64 & 3.55 \\
{[}Ar III{]} & 7135.80 & 14.15 & 14.38 & 15.15 & 15.17 \\
{[}O II{]} & 7319.99 & 1.00 & 0.97 & 1.04 & 1.09 \\
\hline
\end{tabular}\\
\flushleft
\footnotesize{From left to right, the columns represent: The species names (atom/ion), their laboratory wavelengths, the observed nebular line ratios from \cite{aller1981analysis}, and the last three columns are the theoretical line ratios predicted by the photoionization models using the indicated stellar atmosphere approximations.}
\end{table*}

\end{appendix}

\end{document}